\documentclass[a4paper,USenglish,cleveref,autoref,thm-restate,final,pdfa]{lipics-v2021}
\pdfoutput=1

\hideLIPIcs
\nolinenumbers

\title{A Zone-Based Algorithm for Timed~Parity~Games}

\author{Gilles Geeraerts}
       {Université libre de Bruxelles, Belgium}
       {gilles.geeraerts@ulb.be}
       {https://orcid.org/0009-0005-7738-4684}{}

\author{Frédéric Herbreteau}
       {Univ. Bordeaux, CNRS, Bordeaux INP, LaBRI, UMR 5800, 33400 Talence, France}
       {frederic.herbreteau@bordeaux-inp.fr}
       {https://orcid.org/0000-0002-1029-2356}{}

\author{Jean-François Raskin}
       {Université libre de Bruxelles, Belgium}
       {jean-francois.raskin@ulb.be}
       {https://orcid.org/0000-0002-3673-1097}{}

\author{Alexis Reynouard}
       {Université libre de Bruxelles, Belgium}
       {alexis.reynouard@ulb.be}
       {https://orcid.org/0000-0002-7769-3576}{}

\authorrunning{G. Geeraerts, F. Herbreteau, J.-F. Raskin, and A. Reynouard}

\Copyright{Gilles Geeraerts, Frédéric Herbreteau, Jean-François Raskin, and Alexis Reynouard}



\EventEditors{John Q. Open and Joan R. Access}
\EventNoEds{2}
\EventLongTitle{42nd Conference on Very Important Topics (CVIT 2016)}
\EventShortTitle{CVIT 2016}
\EventAcronym{CVIT}
\EventYear{2016}
\EventDate{December 24--27, 2016}
\EventLocation{Little Whinging, United Kingdom}
\EventLogo{}
\SeriesVolume{42}
\ArticleNo{23}

\usepackage{xspace}

\usepackage{algpseudocodex}
\usepackage[plain]{algorithm}

\usepackage{tikz}
\usetikzlibrary{automata, arrows.meta, positioning}
\usepackage[nomessages]{fp}
\usepackage{xcolor}
\definecolor{TargetColor}{rgb}{0.0, 0.43,0.38}    
\definecolor{GoodColor}{rgb}{0.99,0.78,0.07}      
\definecolor{BadColor}{rgb}{0.3,0.18,0.0}         
\definecolor{DisabledColor}{rgb}{0.87,0.88, 0.87} 

\newlength{\tmplen}

\usepackage{pgfplots}
\pgfplotsset{compat=1.18}
\usepackage{alltt}

\usepackage{stmaryrd} 
\usepackage{mathtools} 

\newcommand{\splitatcommas}[1]{%
  \begingroup
  \ifnum\mathcode`,="8000
  \else
    \begingroup\lccode`~=`, \lowercase{\endgroup
      \edef~{\mathchar\the\mathcode`, \penalty0 \noexpand\hspace{0pt plus 1em}}%
    }\mathcode`,="8000
  \fi
  #1%
  \endgroup
} 

\newcommand\uppaal{\textsc{UppAal}\xspace}
\newcommand\uppaaltiga{\textsc{UppAal-TiGa}\xspace}
\newcommand\tga{TGA\xspace}

\newcommand\wlogep{w.l.o.g.}

\newcommand\overax{over actions $A = A_0 \uplus A_1$ and clocks $X$\xspace}
\newcommand\pp{Player~$p$\xspace}
\newcommand\tpl[1]{Player~#1\xspace}

\newcommand{\tst}{\ensuremath{\text{ s.t.\ }}}

\newcommand{\tand}{\ensuremath{\text{ and }}}

\newcommand\Rp{\mathbb R_{\geq0}}
\newcommand\NN{\mathbb N}

\DeclareMathOperator{\fract}{frac} 

\newcommand\lcc{\llbracket} 
\newcommand\rcc{\rrbracket} 
\newcommand\cc[1]{\lcc{#1}\rcc} 

\DeclareMathOperator\last{last}
\DeclareMathOperator\tick{tick}

\newcommand\none{\mathord-}

\DeclareMathOperator\vals{Vals}           

\newcommand\ab[1]{{A_{\bot, #1}}}
\newcommand\deftga{\splitatcommas{(Q, E, C, \urg0, \urg1, \alpha)}}
\newcommand\nametga{G = \deftga}

\newcommand\urg[1]{U_{#1}}

\DeclareMathOperator\cregs{CRegs}         
\DeclareMathOperator\guards{Guards}       

\newcommand\game{\mathcal{G}}
\newcommand\defts{\splitatcommas{(S, M, \tran, \beta)}}
\newcommand\defgame{(\Phi, \wc)}
\newcommand\defgamenamets{\splitatcommas{(\namets, \wc)}}
\newcommand\namets{\Phi = \defts}
\newcommand\namegame{\game = (\Phi, \wc)}

\newcommand\paritye{{\parity_{0}}}          
\newcommand\parityo{{\parity_{1}}}          
\newcommand\tran{\mathord\to}
\newcommand\wc{\mathit{WC}\,}     

\DeclareMathOperator\moves{Moves}         
\DeclareMathOperator\parity{Parity}       

\newcommand\sem{\mathcal{Q}}
\newcommand\sus{\mathcal{S}}
\newcommand\tis{\mathcal{T}}
\newcommand\res{\mathcal{R}}
\newcommand\Eres{\enrich{\res}}
\DeclareMathOperator\blame{bl}

\DeclareMathOperator\lazyblame{lbl}
\DeclareMathOperator\playing{P}

\newbox\enrichbox
\newcommand\enrich[1]{{%
      \mathchoice%
      {{#1^{\scalebox{0.6}{+}}}}%
      {{#1^{\hspace{-1pt}{\raisebox{0.7ex}{\scalebox{0.5}{+}}}}\hspace{-1.1pt}}}%
      {{#1^{\hspace{-0.7pt}\raisebox{0.42ex}{\scalebox{0.45}{+}}}\hspace{-2.2pt}}}%
      {{#1^{\hspace{-0.7pt}\raisebox{0.39ex}{\scalebox{0.42}{+}}}\hspace{-2.2pt}}}%
    }}
\newcommand\EG{\enrich{G}}
\newcommand\defEtga{\splitatcommas{(\enrich Q, \enrich E, \enrich C, \enrich{\urg0}, \enrich{\urg1}, \enrich \alpha)}}
\newcommand\nameEtga{\enrich G = \defEtga}
\newcommand\reset\circlearrowleft

\newcommand\ERG{\Eres(\EG)}

\newcommand\init{\mathit{in}}
\newcommand\out[1]{{\outcome^{#1}}}         
\newcommand\td{\mathit{td}}
\newcommand\zr{Z^\mathrm{r}}                
\newcommand\zz{Z^z}                         

\DeclareMathOperator\outcome{Out}         

\newcommand\tbefore[1]{\ensuremath{#1}\hspace{-1pt}{\rotatebox{3}{\scalebox{0.8}{\ensuremath{\sswarrow}}}}}

\newcommand\zsem[1]{\ensuremath{\cc{#1}}} 
\newcommand\feds{\mathbb{F}}

\DeclareMathOperator\cpre{CPre}           
\DeclareMathOperator\dcpre{DCPre}
\DeclareMathOperator\predt{TPred}

\usepackage{expl3,xparse}
\ExplSyntaxOn
\clist_new:N \l_strategies_clist
\seq_new:N \l_strategies_seq
\newcommand\strategylist[2]{
  \clist_set:Nn \l_strategies_clist {#1}
  \seq_clear:N \l_strategies_seq
  \clist_map_inline:Nn \l_strategies_clist {
    \seq_put_right:Nn \l_strategies_seq {##1\sb#2}
  }
  \int_compare:nTF {\clist_count:N \l_strategies_clist > 1}
    {
      (\seq_use:Nn \l_strategies_seq {\circ})
    }
    {
      \seq_use:Nn \l_strategies_seq {\circ}
    }
}
\ExplSyntaxOff
\NewDocumentCommand\informalstrategylemma{somo}{
  Let $G$ be a \tga.
  If a player has a winning strategy~$\sigma$ in $\res(G)
  \IfNoValueF{#4}{= (\Phi, \wc)}
  $, then one can construct from $\sigma$ a
  \IfNoValueT{#2}{winning}
  #3 strategy~$\sigma'$%
  \IfNoValueF{#2}{ that is winning against opponent's #3 stratgies}%
  \IfBooleanF{#1}{ and the other way around}.
}
\NewDocumentCommand\strategylemma{somm}{
  \IfBooleanTF{#1}%
  {\IfNoValueTF{#2}%
    {\informalstrategylemma*{#4}[L]}%
    {\informalstrategylemma*[R]{#4}[L]}%
  }%
  {\IfNoValueTF{#2}%
    {\informalstrategylemma{#4}[L]}%
    {\informalstrategylemma[R]{#4}[L]}%
  }

  Let $S$ be the set of states of $\Phi$.
  Formally for \tpl0,
  for all states $s_\init \in S$,
  if   $\exists \sigma_0  \in Z_0(\Phi)   :$ $\forall \sigma_1  \in Z_1(\Phi) :$ $\out{s_\init}(\sigma_0,  \sigma_1)  \subseteq \wc$,
  then $\exists \sigma'_0 \in \strategylist{#3}{0}(\Phi) :$ $\forall \sigma'_1 \in
  \IfNoValueTF{#2}
    {Z_1(\Phi)}
    {\strategylist{#3}{1}(\Phi)}
  :$ $\out{s_\init}(\sigma'_0, \sigma'_1) \subseteq \wc$.

  This holds for \tpl1 too.
  Formally,
  if   $\forall \sigma_0  \in Z_0(\Phi) :$ $\exists \sigma_1  \in Z_1(\Phi)   :$ $\out{s_\init}(\sigma_0,  \sigma_1)  \not\subseteq \wc$,
  then $\forall \sigma'_0 \in
  \IfNoValueTF{#2}
    {Z_0(\Phi)}
    {\strategylist{#3}{0}(\Phi)}
  :$ $\exists \sigma'_1 \in \strategylist{#3}{1}(\Phi) :$ $\out{s_\init}(\sigma'_0, \sigma'_1) \not\subseteq \wc$.

  Moreover, for both $p \in \{0, 1\}$, one can construct $\sigma'_p$ from~$\sigma_p$\IfBooleanF{#1}{ and the other way around}.
  Finally, if $G$ is deadlock-free, then there exists complete strategies for both \pp in $\strategylist{#3}{p}(\Phi)$.
}

\begin{document}

\maketitle

\begin{abstract}
  This paper revisits timed games by building upon the semantics introduced in ``The Element of Surprise in Timed Games'' \cite{LAlfaroFaellaHenzingerMajumdarStoelinga03ElementSurpriseTimed.a}.
  We introduce some modifications to this semantics for two primary reasons: firstly, we recognize instances where the original semantics appears counterintuitive in the context of controller synthesis; secondly, we present methods to develop efficient zone-based algorithms.
  Our algorithm successfully addresses timed parity games, and we have implemented it using \uppaal's zone library.
  This prototype effectively demonstrates the feasibility of a zone-based algorithm for parity objectives and a rich semantics for timed interactions between the players.
\end{abstract}

\section{Introduction}\label{sec:introduction}

Timed automata~\cite{RAlurDill94TheoryTimedAutomata.a} are a well-established formalism for modeling real-time systems. Their semantics captures the continuous evolution of time along with discrete transitions, making them particularly suitable for specifying and verifying time-critical behaviors. A crucial aspect in their analysis is the treatment of infinite executions: runs in which time converges are usually deemed unrealistic, as they imply infinitely many actions occurring within a finite amount of time, a phenomenon called \emph{zenoness}.

Timed games extend the model of timed automata to a two-player setting, where one player represents the system under design and the other represents its environment. Defining the semantics of such games is challenging. A central issue is that one player might attempt to win by taking actions increasingly closer in time, effectively blocking the opponent's ability to respond. However, such strategies rely on blocking the progress of time and are thus physically implausible and should not be permitted.

The paper ``The Element of Surprise in Timed Games''~\cite{LAlfaroFaellaHenzingerMajumdarStoelinga03ElementSurpriseTimed.a} proposes an elegant solution to this problem by introducing a timed game model that preserves both the element of surprise (players are allowed to play quickly in order to overtake their adversary), and the need for time divergence. In this model, players independently and simultaneously choose both an action and a delay before executing it. The action with the shortest delay is taken, introducing inherent nondeterminism in the case of equal delays. To ensure physical realism, the semantics prevents a player from winning by stopping time and introduces symmetric winning conditions applicable to all $\omega$-regular objectives. The authors show that the ability to surprise the opponent increases the expressive power of the model and demonstrate that memory strategies may be necessary even for reachability objectives. Although the semantics is shown to be decidable via region-based abstraction, no implementation currently exists. In particular, this semantics requires solving parity games over timed structures, a task for which no practical implementation is available. Existing tools only support simpler objectives such as safety and reachability.

For simpler games like safety and reachability games, the paper~\cite{FCassezDavidFleuryLarsenLime05EfficientFlyAlgorithms.a} presents a fully symbolic zone-based on-the-fly algorithm for solving reachability and safety timed games. This approach is based on a forward exploration of the state space using zones and supports early termination. The algorithm is efficient in practice and was evaluated experimentally, but it relies on a simpler semantics that does not include the surprise mechanism or handle general $\omega$-regular objectives.

Our objective in this work is to provide an efficient implementation of a semantics closely related to that introduced by de Alfaro et al.~\cite{LAlfaroFaellaHenzingerMajumdarStoelinga03ElementSurpriseTimed.a}, which preserves the notion of \emph{surprise} in real-time interactions.
This will be the first efficient implementation of synthesis for parity objective in a timed two-player context.
To this end, we slightly adapt the original semantics with two goals: to make it amenable to a symbolic zone-based implementation; and to handle the controller synthesis problem in a natural way, while retaining the essential features of the original model. We then develop a symbolic algorithm for solving parity games under this expressive timed semantics. Our method builds on ideas from Zielonka's recursive algorithm for parity games on finite graphs~\cite{WZielonka98InfiniteGamesFinitely.a}. To the best of our knowledge, this is the \emph{first implementation} of an algorithm for solving \emph{timed parity games}, marking a significant advance in the analysis and synthesis of real-time systems with complex winning conditions.

\paragraph*{Related works}
Above, we discussed the foundational contributions of de Alfaro et al.~\cite{LAlfaroFaellaHenzingerMajumdarStoelinga03ElementSurpriseTimed.a} and Cassez et al.~\cite{FCassezDavidFleuryLarsenLime05EfficientFlyAlgorithms.a}, which address the semantics and algorithmic analysis of timed games respectively. Beyond these, several studies have explored related challenges in timed games.

Cassez et al.~\cite{DBLP:conf/hybrid/CassezHR02} examine various formulations of control problems for timed and hybrid systems, highlighting the limitations of dense-time controllers and advocating for more implementable discrete-time approaches. Their solution is obtained through a discretization of the timed semantics which is avoided by~\cite{LAlfaroFaellaHenzingerMajumdarStoelinga03ElementSurpriseTimed.a}.
At the initial phase of developing real-time systems, discretization is typically deemed less suitable than continuous time semantics that are more abstract. Brihaye et al.~\cite{DBLP:conf/icalp/BrihayeHPR07} focuses on the minimum-time reachability problem in timed games, presenting methods to compute the least time required for a player to reach a target location against all possible strategies of the opponent. In a case study paper, Cassez et al.~\cite{DBLP:conf/hybrid/CassezJLRR09} demonstrate the application of synthesis tools like \uppaaltiga for the automatic generation of robust and near-optimal controllers in an industrial application.

Addressing the implementability of timed models, De Wulf et al.~\cite{DBLP:journals/fac/WulfDR05} introduce the ``Almost ASAP'' semantics, a relaxation of the traditional ASAP semantics, designed to better reflect the limitations of physical systems, especially when implementing precisely timed strategies on hardware with digital clocks. 

Bouyer et al.~\cite{DBLP:conf/concur/BouyerJM15} investigate the value problem in weighted timed games, proving its undecidability in the general case~\cite{DBLP:conf/formats/BrihayeBR05}, but also providing approximation algorithms for certain subclasses.

Realizability problems for real-time logics have been explored in~\cite{DBLP:conf/formats/DoyenGRR09,DBLP:conf/formats/GiampaoloGRS10}, where it was shown that only specific fragments of MITL are decidable, while the full logic is undecidable.

Finally, Asarin et al.~\cite{DBLP:conf/hybrid/AsarinMP94} propose symbolic techniques for controller synthesis in both discrete and timed systems, aiming to manage large state spaces without relying on exhaustive enumeration, by leveraging logical formulas. This line of work has been further developed through contributions involving the use of zones~\cite{DBLP:conf/rtss/LarsenLPY97}.

\section{Preliminaries}\label{sec:preliminaries}

We begin by defining the arenas on which our games are played. These arenas are essentially Alur-Dill timed automata~\cite{RAlurDill94TheoryTimedAutomata.a}, augmented with both discrete and continuous transitions, which are partitioned into transitions controlled by one player and transitions controlled by the other. We then define a family of functions that associate a game to each arena. These functions correspond to the semantics applied to arenas. A game, in this context, is a transition system equipped with a winning condition for one player; the objective of the other player is adversarial and defined as the complement of the first player's objective.

\paragraph*{Clocks, Valuations, and Guards}\label{sec:clocks-valuations-and-guards}
An \emph{interval} is a nonempty convex subset of the set $\Rp$ of non-negative real numbers.
Intervals can be left-open or left-closed, right-open or right-closed, and bounded or unbounded.
An interval takes one of the following forms: $[a, b]$, $[a, b)$, $[a, \infty)$, $(a, b]$, $(a, b)$, $(a, \infty)$, with  $a, b \in \NN$ and~$a \leq b$.
We write $\emptyset$ for any empty interval.

Let $X$ be a finite set $\{x_1, x_2, \dots, x_n\}$ of variables called clocks.
An \emph{atomic clock constraint} is a formula of the form $x \in I$ where $I$ is an interval.
A \emph{guard} is a Boolean combination of atomic clock constraints.
We denote by $\guards(X)$ the set of all guards on~$X$.
A \emph{valuation} for the clocks in $X$ is a function $v : X \to \Rp$.
To simplify the notation, we sometimes write $v_x$ for $v(x)$.
We write $v \models g$ whenever the valuation $v$ satisfies the guard~$g$.
$\vals(X)$ is the set of all valuations of $X$ and $\mathbf 0_X$ is the valuation that assigns 0 to all clocks of~$X$.
$\top \in \guards(X)$ is called \emph{universal guard} and is satisfied by all valuations.
$\bot \in \guards(X)$ is the \emph{empty guard} and is never satisfied.

For $R \subseteq X$ (resp.\ $x \in X$), we write \emph{$v[R := 0]$} (resp.\ $v[x:=0]$) for the valuation that assigns 0 to all clocks $x \in R$ (resp.\ to $x$), and $v(x')$ to all the other clocks.
For $t \in \Rp$, we write \emph{$v + t$} for the valuation that assigns the value $v(x) + t$ to each clock $x \in X$.
Given two valuations $v \tand v'$, we write $v \bowtie v'$, for $\mathord{\bowtie} \in \{\mathord <, \mathord \leq, \mathord =, \mathord \geq, \mathord >\}$ when $\forall x \in X : v(x) \bowtie v'(x)$.
We can now define the syntax of timed games arenas.

\paragraph*{Timed Game Arena}
We study games played on arenas.
An arena models the system under study.
The definition of the game associated with an arena depends on the semantics under consideration.
The game rules are determined by the specific semantics used.

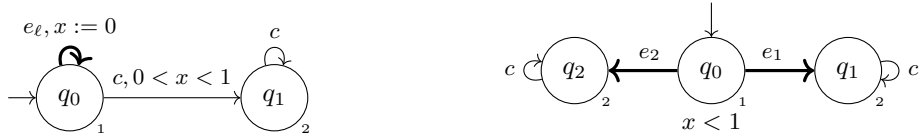
\begin{figure}
  \centering
  \begin{subfigure}[t]{0.5\linewidth}
    \centering
\begin{tikzpicture}[auto, node distance=0.9cm and 1.8cm, align=center, initial text=,
    every label/.style={font=\tiny, inner sep=1pt, circle},scale=.8]
  \node (q2) [state, initial,                label={below right:$1$}] {$q_0$} ;
  \node (q3) [state,          right = of q2, label={below right:$2$}] {$q_1$} ;
  \path [->, every node/.style={font=\footnotesize}]
  (q2) edge [loop above, very thick, looseness=3.6] node {$e_\ell, x:=0$} (q2)
  (q2) edge [thin]                node {$c, 0<x<1$} (q3)
  (q3) edge [loop above, thin, looseness=3.6] node {$c$} (q3)
  ;
\end{tikzpicture}
    \caption{
      A \tga
      from~\cite{LAlfaroFaellaHenzingerMajumdarStoelinga03ElementSurpriseTimed.a}.
      Under the surprise semantics, the Controller can reach $q_1$ from $q_0$ by playing fast enough and retrying as many time as needed.
    }
  \label{fig:2-locations}
  \end{subfigure}
  \hfill
  \begin{subfigure}[t]{0.45\linewidth}
    \centering
\begin{tikzpicture}[auto, node distance=0.9cm and 1.8cm, align=center, initial text=,
    initial where=above,
    every label/.style={font=\tiny, inner sep=1pt, circle}]
  \node (q2) [state,                         label={below right:$2$}] at (-1.8, 0) {$q_2$} ;
  \node (q0) [state, initial,                label={below right:$1$}] at ( 0  , 0) {$q_0$} ;
  \node (q1) [state,                         label={below right:$2$}] at ( 1.8, 0) {$q_1$} ;
  \path [->, every node/.style={font=\footnotesize}]
  (q0) edge [very thick]               node[      pos=0.4] {$e_1$}  (q1)
  (q0) edge [very thick]               node[swap, pos=0.4] {$e_2$}  (q2)
  (q1) edge [loop right, looseness=3.6]          node {$c$}    (q1)
  (q2) edge [loop left , looseness=3.6]          node {$c$}    (q2)
  ;
  \node [style={font=\small}, below] at (q0.south) {$x<1$} ;
\end{tikzpicture}
    \caption{A \tga where the Controller should win (by reaching either $q_1$ or $q_2$).}
    \label{fig:3-locations}
  \end{subfigure}
  \caption{
    Two \tga{}s.
    Thick arrows are Environment's actions.
    The Controller aims to visit locations with even colors, indicated at their bottom right, infinitely often.
  }\label{fig:examples}
\end{figure}

A \emph{timed game arena}, \tga for short, over actions $A = A_0 \uplus A_1$ and clocks $X$ is a tuple $\nametga$ where
$Q$ is a finite set of \emph{locations},
$E \subseteq Q \times A \times \guards(X) \times 2^X \times Q$ is a finite set of labeled \emph{edges},
$C : Q \to \guards(X)$ is an \emph{invariant function},
$\urg p : Q \to \guards(X)$ are the \emph{no-wait functions} of the players $p\in\{0, 1\}$, and
$\alpha : Q \to \{0, 1, \dots, d\}$ is a \emph{coloring function of the locations}.
$\alpha(q)$ is the \emph{color} of~$q$ (sometimes called its \emph{priority}).
The partition of $A$ into $A_0$ and $A_1$ represents the possible actions of two players: \tpl0, which is called the \emph{Controller} and \tpl1, the \emph{Environment}.
Given a \tga $G$, $\guards(G)$ is the set of guards that appear in $G$ (in the image of $C$, $\urg0$, or $\urg1$, or as an edge guard in $E$).
\Cref{fig:2-locations} shows a \tga with one clock $x$ and two locations.
The number next to a location $q$ is its color $\alpha(q)$.
The Controller's actions set is $A_0 = \{ c \}$ and the Environment's actions set (thick arrows) is $A_1= \{ e_\ell \}$.
The guard below a location is its invariant.
Intuitively, the invariant of a location $q$ is a condition on the clocks that must be satisfied for the game to be in $q$.
On this example, all the no-wait functions always return $\bot$ for both players and we do not display them.
Intuitively, the no-wait functions of the players force them to play without a delay when the clocks have certain value.
For example, if $\urg0(q)=x<1$, then \tpl{0} cannot let time elapse in location $q$ whenever the clock $x$ is smaller than $1$.
Those no-wait functions usually do not appear in the classical definitions of timed games, but they will turn out to be useful in the sequel.

To play games on such \tga, we  need  further assumptions.
We extend all \tga $\nametga$ \overax with two extra actions $\bot_0$, $\bot_1$ and an extra clock $z$.
The actions $\bot_p$ will be used in the game setting to mean that \tpl{$p$} will choose to stay in the current location: we assume that each location $q$ bears two self-loops labeled respectively by $\bot_0$ and $\bot_1$.
Formally, we let: $\ab{p} := A_p \cup \{ \bot_p \}$ for $p\in \{0, 1\}$ and $E_\bot := E \cup \big\{ (q, \bot_p, \top, \emptyset, q) \mid \forall q \in Q, p \in \{0, 1\}\big\}$.
When necessary, we call $A_p$ the set of \emph{regular actions} of~$p$.
The clock $z$ will be used to measure elapsed time and is never reset or tested.
We let $X_z = X \cup \{z\}$.
From now on, we assume that all \tga are equipped with $z$, $\bot_0$ and~$\bot_1$.
Note that we do not formally fix an initial location in the definition of a \tga.
We may give one on the figures to help intuition.

\paragraph*{Alur-Dill semantics}
All the semantics we will discuss in \cref{sec:three-semantics} are based on the following definitions.
A \emph{configuration} is a pair $(q, v)\in Q\times \vals(X_z)$.
The \tga can only be in such a configuration if $v$ satisfies the invariant of $q$; and time can elapse, increasing the valuation, as long as the invariant stays satisfied.
From a configuration $(q, v)$, an edge $(q, a, g, R, q')$ can be fired iff the current valuation $v$ satisfies~$g$.
This leads to configuration $(q', v')$, where $v'$ is obtained from $v$ by resetting the clocks in $R$.
Formally, given $(q, v)$, we say that an action $a$ is \emph{enabled} whenever there is $(q, a, g, R, q') \in E$ s.t.\ $v \models g \wedge v[R:=0] \models C(q')$.
We write $(q, v) \xrightharpoonup{\Delta, a} (q', v')$ with $\Delta \in \Rp$ to say that $a$ is enabled in $v + \Delta$, leading from $(q, v + \Delta)$ to $(q', v')$, and that one can wait from $(q, v)$ to $(q, v + \Delta)$ without violating the invariant $C(q)$ nor the no-wait function $\urg p(q)$, with $p \tst a \in \ab{p}$.
Formally, $(q, v) \xrightharpoonup{\Delta, a} (q', v')$ iff:
\begin{multline*}
  \forall \delta \in [0, \Delta] : v + \delta \models C(q) \land  \exists (q, a, g, R, q') \in E_\bot :\\
  \big(v + \Delta \models g \wedge v' = v + \Delta[R:=0] \wedge v' \models C(q')\\
  \land a\in \ab{p} \implies \forall \delta'\in (0, \Delta): v+\delta'\not\models \urg p(q)\big)\text{.}
\end{multline*}

A \emph{deadlock} occurs whenever there is a location-valuation pair $(q, v)$ s.t.\ $C$ limits the time one can stay in $q$ starting from $v$ and there is no action enabled in that time.
Formally, there is a deadlock in $(q, v)$ whenever $\nexists \Delta, a, q', v' : (q, v) \xrightharpoonup{\Delta, a} (q', v')$.
From now on, we assume that our arenas are deadlock-free.

\paragraph*{Clock Regions}
We rely on the usual region construction \cite{RAlurDill94TheoryTimedAutomata.a} that we adapt slightly to account for the special clock $z$.
It is based on a finite equivalence relation over clock valuations of $X_z$.
A clock region is an equivalence class of this relation.
A \emph{region} is a pair $(q, r)$ with $r$ a clock region.
Since there are finitely many regions, and regions form a time-abstract bisimulation, regions form a finite abstraction of the state space that is suitable for our purpose.

Given a \tga~$G$, for each clock $x \in X$, let $c_x$ be the largest constant in $\guards(G)$ that involves $x$, where $c_x = 0$ if $x$ does not occur in any guard, no-wait or invariant of~$G$.
Let $v_1$ and $v_2$ be two clock valuations in $\vals(X_z)$.
We let $X'=\{y\in X\mid v_1(y)<c_y\}\cup\{z\}$ denote the set of clocks $y$ which have a value smaller than $c_y$ in $v_1$ plus the special clock $z$.
Then, $v_1$ and $v_2$ are \emph{clock equivalent}, denoted $v_1 \equiv_G v_2$, iff:
\begin{enumerate}
  \item for all $x \in X$: either
        $\lfloor v_1(x) \rfloor = \lfloor v_2(x) \rfloor$ or
        $\lfloor v_1(x)\rfloor > c_x$ and $\lfloor v_2(x)\rfloor > c_x$;
  \item the ordering of the fractional parts of all clocks in $X'$ is
        the same in $v_1$ and $v_2$;
  \item for all clocks $x\in X'$: $v_1(x)$ is an integer if and only if
        $v_2(x)$ is an integer.
\end{enumerate}
Note that the special case we apply to $z$ can be interpreted as if
$z$ was automatically reset to~$0$ every time it reaches $1$.  Indeed,
the first condition about the integer parts of the clocks does not
constrain $z$, and the two other conditions constrain only the
fractional part of $z$'s value.
Then, a \emph{clock region} is an equivalence class of $\equiv_G$, and
we write \emph{$\cc{v}_G$} for the clock region of the clock
valuation~$v$. It is well-known that $\equiv_G$ is a finite time-abstract bisimulation relation, hence that the number of clock regions is
finite for all \tga~\cite{RAlurDill94TheoryTimedAutomata.a}.

Next, to obtain games, we will associate a \emph{transition system} (whose states and transitions will define the possible game states and moves) and a \emph{winning condition} to each \tga.
We will consider \emph{concurrent games}, where both players choose a move simultaneously and the game state is updated according to the choice of both players.
In \cref{sec:three-semantics}, we will discuss different possible ways to build a transition system for a given \tga.

\paragraph*{Game}
A \emph{(labeled) transition system} over a potentially infinite set $M$ of labels is a tuple $\namets$ where $S$ is a potentially infinite set of states, $\tran \subseteq S \times M \times S$ is a labeled transition relation, and $\beta: S\mapsto\NN$ is a coloring function of the states whose image must be finite.
As usual, we write $s\xrightarrow{m} s'$ instead of $(s, m, s')\in \tran$.

To model concurrent games, the set of labels are of the form $M = M_0 \times M_1$.
The elements of $M_p$ are called the \emph{moves} of \tpl{$p$}.
Intuitively, the game is played \emph{round by round} and each round is characterized by the current game state (assuming a fixed initial state $s_\init \in S$).
When the game is in state $s$, both players concurrently choose respective moves $m_0\in M_0$ and $m_1\in M_1$, and the game moves to any state $s'$ s.t. $s\xrightarrow{m_0, m_1} s'$.
Then, a new round starts and the game continues \textit{ad infinitum}, forming an infinite \emph{play}.

Formally, a \emph{play from $s_\init$} is an infinite sequence of the form $s^0 \big( m_0^i m_1^i s^i \big)_{i \in 1, 2, 3\dots}$ s.t.: $s^0 = s_\init$ and $s^{i-1} \xrightarrow{m_0^i, m_1^i} s^{i}$ for all $i\geq 1$.
A \emph{history} is a sequence as above that is finite.
Let $K(\Phi)$ denote the set of all possible plays on a transition system~$\Phi$
and $H(\Phi)$ be the set of its possible histories.
A \emph{run} is a history or a play.

We use \emph{parity conditions} to define winning plays.
Let $\rho=\big(s^0 \big( m_0^i m_1^i s^i \big)_{i \in 1, 2, 3 \dots}\big)$ be a play
and $\beta$ a coloring function of the states.
We lift the coloring function $\beta$ from states to runs:
$\beta(\rho)=\big(\beta(s^i)\big)_{i \in 0, 1, 2\dots}$.
We denote $\inf(\beta(\rho))$ the set of colors that occur infinitely often in $\rho$.
Then, we write $\rho \models \paritye(\beta)$ iff $\max(\inf(\beta(\rho)))$ is even and $\rho \models \parityo(\beta)$ iff $\max(\inf(\beta(\rho)))$ is odd.

A \emph{game} is obtained by associating a winning condition to a transition system.
A game is thus a pair $\game = (\Phi, \wc)$ where $\Phi$ is a transition system over a set $M_0 \times M_1$ and $\wc \subseteq K(\Phi)$ a set of \tpl{0}'s \emph{winning} plays.
In games, players can play according to a strategy.
A \pp's \emph{strategy} is a function $\sigma_p : H(\Phi) \to M_p$ that associates a \pp's move to all histories.
We denote by $Z_p(\Phi)$ the set of all strategies of \tpl{$p$} in~$\Phi$.
A run $\rho = s^0(m_0^i m_1^i s^i)_{i \in 1,2,\ldots}$ is played according strategies $\sigma_0$ and $\sigma_1$ if for every $i \ge 1$, $m_0^i = \sigma_0(h_i)$ and $m_1^i = \sigma_1(h_i)$ where $h_i$ is the prefix of $\rho$ up to $s^{i-1}$.
Then, given two strategies $\sigma_0$ and $\sigma_1$, the \emph{outcome} $\out{s_\init}(\sigma_0, \sigma_1)$ is the set of all plays from $s_\init$ played according to $\sigma_0$ and $\sigma_1$.
A strategy $\sigma^\star$ of the Controller (\tpl{0}) is \emph{winning from $s_\init$} iff $\out{s_\init}(\sigma^\star, \sigma_1)\subseteq \wc$ \textbf{for all} Environment's strategy~$\sigma_1$.

Finally, we assume that all games $\game=(\Phi, \wc)$ are
\emph{well-defined}, i.e.\
(1) 
$\Phi$ is deadlock-free
and
(2) 
in all states, the moves available to a player do not
depend on the move played by the other player.
Formally,
for any state $s \in S$, let
$\moves_0(s) = \{m_0 \in M_0 \, | \, \exists m_1 \in M_1: \, \exists
s' \in S \text{ s.t. } s \xrightarrow{m_0,m_1} s'\}$. We define
$\moves_1(s)$ symmetrically for \tpl1.  Then, for all $s\in S$, there
are $m_0, m_1, s'$ s.t. $s \xrightarrow{m_0, m_1} s'$, \emph{and} for
all $s\in S$, for all $m_0 \in \moves_0(s)$, for all $m_1$ in
$\moves_1(s)$: there is $s'$ s.t. $s \xrightarrow{m_0, m_1} s'$.

Deciding how to build a game from an arena is not a trivial problem, and we devote Section~\ref{sec:three-semantics} to discuss several possibilities.

\section{Three Semantics for Timed Games}\label{sec:three-semantics}

In the present section, we discuss three possible ways to associate a game $\namegame$ to a \tga $G$, thereby providing three possible definitions of game semantics in a unified framework.
This framework constitutes the first contribution of the paper.
We first recall the ``classical'' semantics of timed games, and the ``surprise'' semantics of \cite{LAlfaroFaellaHenzingerMajumdarStoelinga03ElementSurpriseTimed.a}.
Then, as a second contribution we introduce a third, novel, semantics that refines and improves the ``surprise'' semantics.
This refinement results in a semantics which is more natural, and that allows for efficient winning strategy computation.
In Section~\ref{sec:solv-synth-probl}, we will introduce a zone-based symbolic algorithm to compute winning strategies in this semantics.

Let us fix a \tga $\nametga$ \overax.
We show how to build a game $\defgame$.
Since we consider timed games, the states of the transition system $\Phi$ are configurations $(q, v)$ of the \tga $G$, and the moves of \tpl{$p$} are pairs $(\Delta, a)\in(\Rp\times \ab{p})$.
Intuitively, each round of the game starts in a state $(q, v)$.
Both players propose concurrently a delay $\Delta$ and an action $a$.
The player with the shortest delay wins the round and his proposal $(\Delta, a)$ is played in $G$: we let $\Delta$ time units (t.u.) elapse, reaching $(q, v+\Delta)$; then follow an $a$-labeled edge.
The special actions $\bot_0$ and $\bot_1$ mean that the player only wants to wait without following any edge.

Plays are now of the form $\rho=(q^0, v^0) (\Delta_0^1, a_0^1) (\Delta_1^1, a_1^1) (q^1, v^1)\ldots$.
We say that such a play is \emph{time-divergent} iff the special clock $z$ cannot be bounded along $v^0, v^1, v^2, \ldots$, i.e. $\forall t \in \Rp : \exists i \in \NN : v^i_z >t$.
We write $\rho \models \td$ if $\rho$ is time-divergent.

We formally define a semantics $\sem$ as a function that associates a game $\namegame$ to a \tga $G$.
We now present the three semantics mentioned above.
Given a \tga $G$, $\tis(G)$ is the timed game classically associated to $G$ \cite{RAlurDill94TheoryTimedAutomata.a}, $\sus(G)$ is the game from \cite{LAlfaroFaellaHenzingerMajumdarStoelinga03ElementSurpriseTimed.a} which accounts for the surprise effect, and $\res(G)$ is a new refinement of $\sus(G)$.
These three semantics are \emph{sound}, i.e., given a deadlock-free arena, they give a game with a well-defined transition system.

\paragraph*{Classical Timed Semantics $\tis$}
The ``classical'' timed semantics associated to $G$ is the game $\tis(G)$.
In this settings, players can only play moves of the form $(0, a)$, meaning ``play $a$ immediately''; or $(\Delta, \bot)$, meaning ``wait for $\Delta$ t.u.''.
For example, consider the \tga in \cref{fig:2-locations}.
The objective of the Controller is to reach $q_1$ from $q_0$.
To do so, he needs to wait in $q_0$ until $x$ becomes non-null, because of the guard from $q_0$ to $q_1$.
However, the Environment can always overtake him by playing $(0, e_\ell)$ that resets the clock.
The Environment can keep playing that way, so, with the classical semantics, there is no winning strategy for the Controller in this game.

Formally, $\tis(G)=\defgamenamets$ is defined as follows.
The state set is $S= Q \times \vals(X_z)$.
The possible moves of \pp are $M_p=(\Rp \times \{ \bot_p \}) \cup (0 \times A_p)$.
The definition of the transition relation $\tran$ follows the above intuition: $(q, v) \xrightarrow{(\Delta_0, a_0), (\Delta_1, a_1)} (q', v')$ iff there is $j\in{0, 1}$ s.t.
$\Delta_j \leq \Delta_{1-j}$ and
$q'=q_j$, $v'=v_j$ with\footnote{Recall that $\rightharpoonup$ is defined above under the paragraph about the Alur-Dill semantics.}
$(q, v) \xrightharpoonup{\Delta_0, a_0} (q_0, v_0)$ and
$(q, v) \xrightharpoonup{\Delta_1, a_1} (q_1, v_1)$.
For the winning condition, we lift the coloring function $\alpha$ on the locations of $G$ to the states of $\Phi$, letting $\beta(q, v)=\alpha(q)$ for all states $(q, v)$.
Then, $\wc$ is defined as a parity condition: $\wc= \{ \rho \in K(\Phi) \mid \rho \models \paritye(\beta) \}$.

\paragraph*{TEOS Surprise Semantics $\sus$}\label{sec:classical-surprise-semantics}
The winning strategy of the Environment described above (always
playing $(0,e_\ell)$ in $q_0$) is problematic since any play in its
outcome is \emph{not} time-divergent.
This is regarded as a flaw in
the model, since playing an infinite number of actions in a finite
time is not realistic.
This observation has prompted the introduction
of the ``surprise'' semantics by de Alfaro et al.~\cite{LAlfaroFaellaHenzingerMajumdarStoelinga03ElementSurpriseTimed.a}.
We called this semantics the ``TEOS semantics'', from the title of the paper, and denote it $\sus$.
It differs from $\tis$ by the definition of the player moves and of the set of winning plays.
In the TEOS semantics, we let $M_p = (\Rp \times \ab{p})$.
Thus, players can propose to wait \emph{and, immediately after}, play an action.
This is where the ``surprise'' stems from: it gives players the ability to
\emph{act quickly} and to take their opponent by surprise by playing
an action after an arbitrarily short delay.
However, plays where the time does not diverge should be avoided, so the game $\sus(G)$ introduces a \emph{blame function} $\blame$ to identify the player blocking the time.
If the Controller is to blame in a play where time does not diverge, he looses the game even if the play satisfies his parity objective.

Formally, we let:
$\blame: S \times M_0 \times M_1 \times S \to \{0, 1\}$ be s.t.
$\blame(s,m_0,m_1, s')=0$ iff $s \xrightharpoonup{m_0} s'$.
Then, $\blame$ is lifted to runs by recording who is playing at each round:
$\blame\big(s^0 \big( m_0^i m_1^i s^i \big)_{i \in 1, 2, 3\ldots}\big) =\big( \blame(s^{i-1}, m_0^i, m_1^i, s^i) \big)_{i \in 1, 2, 3\ldots}$.
\tpl{$p$} will be to blame in play $\rho$ iff $p$ occurs infinitely often in $\blame(\rho)$.
The Controller then wins iff it satisfies his parity objective on time-divergent plays \emph{and} is not to blame if the time converges.
Formally, for $p\in\{0,1\}$, we write
$\rho \models \blame_p$ iff $p \in \inf(\blame(\rho))$, and let:
$
  \wc = \{ \rho\mid
  \big( \rho \models \td \wedge \rho \models \paritye(\beta) \big)
  \vee
  \big( \rho \not\models \td \wedge \rho \not\models \blame_0 \big) \}
$.
The other components of $\sus(G)$ are the same as for~$\tis(G)$.

Consider again \cref{fig:2-locations}, and the Environment's strategy
$\sigma^\star_1$ that consists in always playing $(0, e_\ell)$ in
$q_0$.  In the TEOS semantics, the Controller has a strategy
$\sigma_0^\star$ to win against $\sigma^\star_1$ that consists in
always playing $(\varepsilon, c)$, for $0<\varepsilon<1$.  In the
outcome of those strategies, the time converges, but the Controller is
not to blame, so he wins those plays although his parity objective is
not fulfilled.
However, such a $\sigma_0^\star$ is \emph{not} winning for the
Controller against \emph{all} strategies of the Environment, who can
always play $(\frac\varepsilon2,e_\ell)$. Yet, the Controller now has a
winning strategy $\sigma_{2^{-i}}$ that consists in playing
$(2^{-i}, c)$ at the $i$-th round until $q_1$ is reached, and
then playing $(1,c)$ for ever. Against $\sigma_{2^{-i}}$, the only way
for the Environment to avoid reaching $q_1$ is to always play smaller
delays to win each round, such as $(2^{-i-1}, c)$ at the $i$-th
round. Then, if $q_1$ is not reached, time converges and the
Environment is to blame, so the Controller always wins, even if $q_1$
is never reached. This illustrates the \emph{ability to act quickly} of
the surprise semantics.

\paragraph*{Refined Surprise Semantics $\res$}
Our last semantics is a novel refinement of $\sus$. The main idea with
this new semantics is that we  allow the Controller \emph{not to
  act} in the states of the system where he has no (discrete) action
at his disposal. Consider an example where the Controller needs to
observe the system and act after the environment has chosen between
two possible actions. Such a situation is modeled in
\cref{fig:3-locations}, where the Controller observes the system in
state $q_0$ (where he has no action) and waits for the Environment to
play either $e_1$ or $e_2$, that will arrive within one time
unit.   In the
$\sus$ semantics, when in $q_0$, the Controller can only propose to
play a sequence of the form
$(\Delta_0,\bot),(\Delta_1,\bot),\ldots,(\Delta_i,\bot),\ldots$ where
$\sum_{i\geq 0}\Delta_i < 1$, in order to satisfy the invariant. Then,
the Environment can always play $(t,e_1)$ with $t\geq t_i$ to let the
Controller play at all rounds.  In such plays, the time converges with
the blame on the Controller who loses the game.
Thus, surprisingly, the Environment wins by doing nothing, forcing the
Controller to block time.  Indeed, the TEOS semantics makes no
distinction between a player who waits because he has no other choice
and a player who voluntarily blocks the time. This prompts the first
refinement of our semantics that consists in introducing a new move,
denoted ``$\none$'' and meaning that the player does not play
anything.

Formally, we extend the set of possible moves of \pp to: $M_p = (\Rp \times \ab{p}) \cup \{ \none \}$.
We define an ancillary predicate $\playing_p(q, v)$ which is true iff \tpl{$p$} can propose a regular action from $(q, v)$, i.e. iff there are $\Delta\in \Rp$, $a\in A_p$, $q'\in Q$ and $v'\in \vals(X_z)$ s.t.  $(q, v) \xrightharpoonup{\Delta, a} (q', v')$.
Observe that $\neg \playing_p(q, v)$ implies $\playing_{1-p}(q, v)$ because we have assumed that $G$ is deadlock-free.
Then, we adapt the transition relation as follows.
\begin{itemize}
  \item $(q, v)\xrightarrow{(\Delta_0, a_0), (\Delta_1, a_1) }(q', v')$
        iff both players can play:
        $\playing_0(q, v) \wedge \playing_1(q, v)$ and $\exists j\in{0, 1}$ s.t.\ $\Delta_j \leq \Delta_{1-j}$ and $q'=q_j$, $v'=v_j$ with $(q, v) \xrightharpoonup{\Delta_0, a_0} (q_0, v_0)$ and $(q, v) \xrightharpoonup{\Delta_1, a_1} (q_1, v_1)$.
  \item $(q, v)\xrightarrow{(\Delta_0, a_0), \none}(q', v')$
        iff only \tpl{0} can play:
        $\neg \playing_1(q, v)$ and $(q, v) \xrightharpoonup{\Delta_0, a_0} (q', v')$.
  \item $(q, v)\xrightarrow{\none, (\Delta_1, a_1) }(q', v')$
        iff only \tpl{1} can play:
        $\neg\playing_0(q, v)$ and $(q, v) \xrightharpoonup{\Delta_1, a_1} (q', v')$.
\end{itemize}
With this change, the Controller no longer needs to propose delays from $q_0$ in \cref{fig:3-locations}.
Instead, in $q_0$, the Controller does not play, and the Environment must play $e_1$ or $e_2$ to model an event; otherwise, the Environment loses by blocking time.

In the game $\res(G)$, we also adapt the \emph{blame} predicate.
In the $\sus$ semantics, the blame only records which Player played last, no matter which delay he has played.
We introduce a \emph{lazy blame} policy, which differs from the classic blame in that a player who has played \emph{a wait $(\Delta, \bot_p)$ long enough to leave the current clock region} is not blamed.
While this change might sound inconsequential, it is
important to obtain an efficient zone-based algorithm later in the
paper.
Formally, we let
$\lazyblame : S \times M_0 \times M_1 \times S \to \{\none, 0, 1\}$ be
s.t. $\lazyblame((q, v), m_0, m_1, (q', v'))=\none$ if $q'=q$ and
$v \not\equiv_G v'$; \emph{otherwise}
$\lazyblame((q, v), m_0, m_1, (q', v'))=\blame((q, v), m_0, m_1, (q', v'))$.
We lift $\lazyblame$ to runs like we did for $\blame$ and write $\rho \models \lazyblame_p$ iff $p \in \inf(\lazyblame(\rho))$.
One can show that it is feasible to replace the blame ($\none$), assigned by $\lazyblame$ s.t.\ no one is blamed, with a repetition of the previous blame.
This allows reducing the state space.
In the next section, we focus on this $\res$ semantics and show
how to solve the synthesis problem in this setting.

\section{Solving the Synthesis Problem On \texorpdfstring{$\res$}{ℛ}}
\label{sec:solv-synth-probl}

Given a well-defined game $\game$ and an initial state $s_\init$, the
\emph{synthesis problem} asks to compute, if possible, a winning
strategy from $s_\init$ for the Controller.  Solving the synthesis
problem on $\res$ is to solve the synthesis problem for all game
$\game = \res(G)$ for all TGA $G$, and all initial configuration
$(q, \mathbf 0_X)$ for all location $q$ of $G$.

In this section, we present an algorithm to solve the synthesis
problem on $\res$. The main technicality to achieve this is to define
an \emph{augmented}\footnote{As we will see later, `augmented' means
  that each location contains two extra discrete pieces of
  information: a color and the identifier of a Player. } \tga $\EG$,
and associate to it an appropriate semantics\footnote{$\enrich\res$ is
  the same as $\res$ but adapted to the new syntax of $\EG$. } to
obtain a game $\ERG$ on which the synthesis problem is easier to
solve. This transformation moves just enough complexity from the
$\res$ semantics to the \tga $\EG$: for instance, in $\EG$, the
locations are augmented with information about the past.  This allow
us to obtain a game with a \emph{pure parity objective}, i.e., a game
whose objective is of the form
$\{\rho\mid \rho\models \paritye(\alpha)\}$ (while in $\res(G)$ the
objective must also account for the blame).

Equiped with $\ERG$, we could imagine a simple algorithm to solve the
synthesis problem. It would work as follows. First, remove the
concurrency of the game by letting the Controller make his choice
first, then letting the Environment play against this move. This
reduction is valid since we are trying to compute a winning strategy
of the Controller against \emph{any} strategy of the environment.  One
can also let the environment lift the non-determinism at this stage by
letting him choose from whom the action will be played when the two
have chosen the same delay.  This is valid since we want a
Controller's strategy s.t.\ \emph{all the plays} in all the outcome
sets are winning.  Second, use the classical region construction to
obtain a finite game. The resulting game is a finite, turn-based game
played on graph with a parity objective, i.e.\ a \emph{parity game}.
Such games are well-studied \cite{WZielonka98InfiniteGamesFinitely.a}
and enjoy positional strategies that can be computed by the classical
Zielonka algorithm.

While this naive algorithm would be correct, we argue it would be
inefficient and would not be benefit from the zone data-structure
classically used to effectively handle set of regions.  We argue in
\cref{sec:algorithmic-properties} that the fixed point computation
would end up splitting many zones into individual regions, which would
obliterate the advantage of using zones instead of regions. Instead,
we propose a different algorithm (see \cref{algo:zielonka}), which is
inspired from the Zielonka algorithm, but works directly on the
structure of $\EG$ to avoid these pitfalls. In particular, we prove
that, in $\EG$ (under semantics $\enrich\res$), one can solve synthesis by
considering a restricted class of strategies only
(\cref{sec:strategies-to-play,sec:algorithmic-properties}).  We
exploit these properties to obtain an efficient zone-based
algorithm. Finally, from the winning strategies computed in $\ERG$, we
can extract winning strategies on $\res(G)$.

Throughout this section, we fix a \tga $\nametga$ over
actions $A = A_0 \uplus A_1$, with set of clocks~$X_z$, and maximum
color $d$.  We also fix $\res(G) = \namegame$.
We extend the notion of region to states $(q,v)$. We say that two
states $s=(q, v)$ and $s'=(q', v')$ are \emph{region equivalent}, or
\emph{in the same region}, written \emph{$s \equiv s'$}, whenever
$q=q'$ and $v\equiv v'$.  Then, a \emph{region}\label{sec:def-regions}
is an equivalence class of states wrt~$\equiv$.  We write
\emph{$\cc s_G$} for the region containing~$s$. Those notions extend
naturally to plays and histories of the same length

\paragraph*{Zones and zone federations}
To manipulate convex sets of clock valuations, we rely on the
classical zone data-structure.  An \emph{atom} is an inequality of the
form $x\bowtie c$ or $x-y\bowtie c$, where $x, y\in X_z$ are clocks of
$G$;
$\mathord{\bowtie}\in\{\mathord <, \mathord \leq, \mathord =, \mathord
\geq, \mathord >\}$; and $c$ is an integer.  Then, a \emph{zone} is a
finite conjunction of atoms.  For example, $(x\leq 5) \wedge (y-x\geq 3)$.
Given a zone $Z$, its semantics $\zsem{Z}\subseteq\vals(X_z)$ is the
set of all valuations that satisfy all the atoms in $Z$, using the
usual semantics for the comparison operators.  
To manipulate non-convex sets of valuations, we rely on \emph{zone
  federations}
\cite{FCassezDavidFleuryLarsenLime05EfficientFlyAlgorithms.a}, which
are finite sets of zones. The semantics of a federation
$F=\{Z_1,Z_2,\ldots,Z_n\}$ is the union
$\zsem{F}=\zsem{Z_1}\cup\zsem{Z_2}\cup\cdots\cup\zsem{Z_n}$. Note that
guards can be encoded as federations.

Given a federation $F$, one can compute efficiently
\cite{FCassezDavidFleuryLarsenLime05EfficientFlyAlgorithms.a} a
federation $\tbefore{F}$ representing all the valuations that can
reach $F$ by letting time elapse, i.e.
$\zsem{\tbefore{F}}=\{v\mid \exists \Delta\in\Rp: v+\Delta\in
\zsem{F}\}$.  Similarly, one can compute $Z[X:=0]$
s.t. $\zsem{Z[X:=0]}=\zsem{Z}[X:=0]$ for a set of clocks $X$ to
reset. Given federations $F_1$ and $F_2$, one can also compute
$F_1\vee F_2$, $F_1\wedge F_2$ and $F_1 - F_2$ s.t.
$\zsem{F_1\vee F_2}=\zsem{F_1}\cup\zsem{F_1}$,
$\zsem{F_1\wedge F_2}=\zsem{F_1}\cap\zsem{F_1}$ and
$\zsem{F_1 - F_2}=\zsem{F_1}\setminus\zsem{F_2}$.  Federations also
allow for inclusion and intersection checks.  Thus, computing all the
valuations of $F$ that satisfy some guard or invariant $g$ encoded
into a federation $F_g$ amounts to computing $F\wedge F_g$.  We write
$\top$ for the universal federation: $\cc \top = \vals(X_z)$; and
$\bot$ for the empty: $\cc \bot = \emptyset$.  For convenience, we
identify a zone to the singleton federation containing it.  Given a
set of clocks $X$, $\feds(X)$ is the set of all the zone federations
one can create with the clocks of $X$.  Federations naturally allow
for the encoding of infinite set of states as finite set of pairs
(location, federation):
$\zsem{ \{(q_1, F_1), (q_2, F_2), \dots \} } = \{ (q, v) \in Q \times
\vals \mid \exists i : q = q_i \wedge v \in \zsem{F_i} \}$.

\subsection{Reduction to a Parity Game}\label{sec:reduction-to-parity}

We apply ideas from \cite{LAlfaroFaellaHenzingerMajumdarStoelinga03ElementSurpriseTimed.a} to reduce the game $\res(G) = \defgame$ to a parity game $\ERG$.
First, we prove that for all strategies $\sigma$ in $\res(G)$, there
exists a strategy $\sigma'$ that stops at each integer bound of $z$
(called a \emph{tick}) and is winning if $\sigma$ is winning.
A \emph{$z$-stepped strategy} is a strategy that always plays a delay
sufficiently small to prevent $z$ from going beyond its next integer
bound.  Formally, a strategy $\sigma$ is a $z$-stepped \pp's strategy
when, for all histories $h$ ending in a state $(q, v)$, we have
$\sigma(h) = (\Delta, a)$ with
$v_z + \Delta \leq \lfloor v_z \rfloor + 1$.  Given a TS $\Phi$,
$\zz_p(\Phi)$ is the set of all $z$-stepped \pp strategies in~$\Phi$. Then:

\begin{lemma}[$z$-stepped Strategies Are Sufficient]
  \label{theo:z-stepped-strat}
  \informalstrategylemma*{$z$-stepped}
\end{lemma}
Intuitively, $\sigma'$ is built to mimic $\sigma$ closely. For
example, from a state where $z = 0.5$, $\sigma$ plays $(1, a)$, then
there is a winning strategy $\sigma'$ that first plays $(0.5, \bot_p)$
to reach $z = 1$, then, if the proposed wait actually happened, plays
$(0.5, a)$ to play $a$ in the valuation chosen by $\sigma$.

\Cref{theo:z-stepped-strat} allows us to incorporate in the arena a
modulo 1 behavior of $z$.  Intuitively, in $\nameEtga$, $z$ evolves in
$[0, 1)$, and is always reset by the only possible action $\reset$
whenever it reaches $1$. On all locations $q$, the invariant becomes
$C(q) \wedge z \leq 1$ and we add self-loops with action $\reset$ that
reset $z$ when $z=1$. We also check that $z < 1$ on all
other edges.

Next, we need to encode the blaming mechanism in the colors of the
parity game.  Whenever there is a tick of $z$, we see color $2+c$,
where $c$ is the largest color seen since the last tick (in the
original game). On all the states where no tick has occurred, we see
color $1-b$, where $b$ is the player to be blamed. So, if the
Controller is to be blamed in a time-convergent play, color $1$ will
be seen infinitely often and the Controller will lose.

More formally, we have
$\enrich Q=\{ (q, c, b) \in Q \times \{0, \dots, d\} \times \{0, 1,
\none\} \mid c \geq \alpha(q) \}$. Each location $\EG$, is thus of the
form $(q, c,b)$, where $c$ is the largest color seen since the last
tick, and $b$ is the blamed player. The set of actions is
$A \cup \{ \reset \}$. No-wait and coloring functions are lifted to
$G$'s locations: for $f \in\{\urg0, \urg1, \alpha\}$,
$\enrich{f}((q, c, b)) = f(q)$. Finally, the set of edges is $\enrich{E} = \splitatcommas{\{ ( (q,c,b), \reset, z = 1, \{z\}, (q,\alpha(q),b) ) \mid (q,c,b) \in \enrich{Q} \}}
\cup
\{ ( (q, c, b), a, g \wedge z < 1, R, (q', \max(c, \alpha(q')), b') )
\mid (q,a,g,R,q') \in E, b' \tst a \in \ab{b'} \}$.

Now that we have defined the arena $\enrich G$, we need to associate
to it a semantics $\ERG$. To this end, we define $\Eres$ as $\res$
except that: (1) players cannot play $(\Delta, \reset)$ with
$\Delta > 0$ and (2) the coloring function is $\enrich \beta$ defined
as $\enrich \beta((q,c,b), v)=1-b$ if $v_z < 1$, and
$\enrich \beta((q,c,b), v)=c+2$ if $v_z=1$.  Hence, $\ERG$ is a game
with a pure parity condition, where the colors have been adapted to
let any player that blocks the time lose the game. Further, players
cannot play $(\Delta, \reset)$ with $\Delta > 0$, which preserves the
winner of the game by \cref{theo:z-stepped-strat}. Since $\ERG$ is
built from $\res$, it also incorporates the \emph{refined surprise
  rule} (that removes the need for a player without an enabled regular
action to play), and the \emph{lazy blame policy} (that does not
assign blame to a player who has waited long enough to exit the
current clock region).

\subsection{Further restricting the strategies}\label{sec:strategies-to-play}

Let us now show that the \emph{lazy blame policy} and the
\emph{refined surprise rule} allow us to further restrict the family
of strategies we can consider to solve synthesis on $\res$.

First, we fix some vocabulary.  We say that, in a given state, a delay
$\delta$ is \emph{short} if we stay in the same region after waiting $\delta$
t.u..  Otherwise, a delay $\Delta$ is long (we use $\delta$ to emphasize that
a delay is short).  A short wait (long wait, short action, long action) is a
move of the form $(\delta, \bot_p)$ (resp.\ $(\Delta, \bot_p)$, $(\delta,a)$,
$(\Delta,a)$). Finally, a wait or action is \emph{immediate} when $\delta=0$,
and \emph{delayed} when $\Delta>0$.
Next, $\tick$ is a function that associates to all runs
$s^0 m^1_0 m^1_1 s^1 \dots$ a sequence of Booleans $(t^i)_{i>0}$
s.t.\ $t^i=\top$ iff $s^i$ is a tick.  Formally,
$\tick((q, v), (q', v')) = \top$ iff
$(\fract(v'_z) = 0) \wedge (v_z \neq v'_z)$; $\tick(s^0) = \top$; and
$\tick(\dots s^{i-1}(m^i_0, m^i_1) s^{i}) = \tick(s^{i-1}, s^i)$ for
all $i\geq 1$.  Then, a strategy $\sigma$ is a \emph{region
  strategy}\label{sec:region-strategies} whenever, for all pairs of history
$(\rho_1, \rho_2) \tst \rho_1 \equiv \rho_2 \tand \tick(\rho_1) =
\tick(\rho_2)$, we have
$\last(\rho_1) \xrightharpoonup{\sigma(\rho_1)} s_1$ and
$\last(\rho_2) \xrightharpoonup{\sigma(\rho_2)} s_2$ with
$s_1 \equiv s_2$.  We denote the set of region strategies for \pp as
\emph{$\zr_p(G)$}. Then:
\begin{lemma}[Region Strategies Are Sufficient]\label{theo:region-strat}
  \informalstrategylemma*{region $z$-stepped}
\end{lemma}

The previous lemma implies that if a strategy is winning, then any
strategy that differs only in the chosen delays, but not enough to
change the regions or ticks, is also winning. Indeed, by
\cref{theo:region-strat}, any short delayed move can be replaced
by an immediate one:

\begin{corollary}\label{cor:no-short-delayed}\label{theo:no-short-non-immediate-moves}
  If there is a \pp's winning strategy $\sigma$, then there is a \pp's
  winning strategy $\sigma'$ that never plays short delayed moves.
\end{corollary}

Next, we show that \emph{split strategies} are
sufficient. Intuitively, split strategies always propose either an
immediate action or a long wait.  Formally, a \pp's strategy $\sigma$
is a \emph{split strategy} if for all history $h$ 
we have $\sigma(h) = (\Delta, a)$ with
$(a \neq \bot_p \implies \Delta = 0) \wedge (\Delta > 0 \implies a =
\bot_p)$.

We claim that, under the $\res$ semantics, for every winning strategy,
there exists a split winning strategy. This might sound surprising
since the ability to play the combination of a delay and a move seems
to be at the heart of the notion of ``surprise'' in timed games. Let us
explain why, \emph{under our new $\res$ semantics}, it is not the
case. First observe that, thanks to
Corollary~\ref{cor:no-short-delayed}, players do not need to play
\emph{short actions} with a non-null delay. Next, we need to explain
why playing \emph{long actions} are not necessary either thanks to our
new definition of the blame.
To illustrate this, we consider the \tga on
\cref{fig:2-locations}. Recall that the controller has a winning
strategy $\sigma_{2^{-i}}$ that consists in playing $(2^{-i}, c)$ at
the $i$-th round until $q_1$ is reached. That is, the Environment may
prevent the Controller to play $c$, but the Controller is able to
retry as long as needed. Consider now the split strategy
$\sigma_{2^{-i}}'$ that plays $(2^{-i}, \bot_0)$ while on $(q_0, v)$
with $v_x = 0$, and $(0, c)$ whenever in $(q_0, v)$ with $v_x > 0$. We
argue that this strategy is losing with $\sus$'s blame policy, but
winning in $\res$. Under $\sus$, the Environment wins by playing
$(1, \bot_1)$ in $(q_0, v_x = 0)$, and $(0, e_\ell)$ in
$(q_0, v_x > 0)$. Indeed, consider the following time-converging play
in the outcome of $\sigma_{2^{-i}}$ and $\sigma_{2^{-i}}'$:
$(q_0, v_x = 0) \xrightarrow{(0.5, \bot_0), (1, \bot_1)} (q_0, v_x =
0.5) \xrightarrow{(0, c), (0, e_\ell)} (q_0, v_x = 0)
\xrightarrow{(0.125, \bot_0), (1,\bot_1)} (q_0, v_x = 0.125) \cdots$,
where the non-determinism is always resolved for the Environment.
In this play, every other move played by the Controller is a long wait (because we leave the region where the clock $x$ is equal to~0).
Under $\sus$'s blame policy, the Controller is blamed for the time-convergence and looses the game.
Now, consider $\res$'s lazy blame policy which does not blame players that \emph{play a long wait}.
Under this semantics, $\sigma_{2^{-i}}'$ is now a winning \emph{split-strategy}.
This example shows why, in the $\res$ semantics, we can restrict ourselves to such strategies:
\begin{lemma}[Split Strategies Are Sufficient]\label{theo:split-strat}
  \informalstrategylemma*{split region $z$-stepped}
\end{lemma}


\subsection{Synthesis Algorithm}\label{sec:algorithm}

We propose \cref{algo:zielonka} to solve the synthesis problem on all
\tga, under semantics $\res$.  Inspired by Zielonka algorithm, our
algorithm is recursive, it computes the winning partition
$W=(W_0, W_1)$ of the state space of $\ERG$, and thanks to a trivial
modification we obtain the winning Controller's strategy from each
state of $W_0$.  By construction, from all winning strategies in
$\ERG$ we can construct a winning strategy in~$\res(G)$.  We first
describe the objects that the algorithm operates on.  We then explain
the algorithm step by step, providing intuition for the proof of
correctness.

\noindent
\begin{minipage}{8cm}
  \begin{algorithm}[H]
    \begin{algorithmic}[1]
      \Function{Solve}{$G$}
      \State $k \gets \max(\enrich{\beta}(S))$
      \State $j \gets k \mod 2$ ; $i \gets 1 - j$
      \State $T_j \!\gets\! \Call{Attractor}{G, j, \cc{k}_{\enrich{\beta}}}$
      \If{$T_j = S$}
      \State $W_i \gets \emptyset, W_j \gets S$
      \State \Return $W$
      \EndIf
      \State $W' \gets$ \Call{Solve}{$G \setminus_j T_j$}
      \If{$W'_i = \emptyset$}
      \State $W_i \gets \emptyset, W_j \gets S$
      \State \Return $W$
      \EndIf
      \State $T_i \gets \Call{Attractor}{G, i, W'_i}$
      \State $W'' \gets \Call{Solve}{G \setminus_i T_i}$
      \State $W_i \gets S \setminus W''_j, W_j \gets W''_j$
      \State \Return $W$
      \EndFunction
    \end{algorithmic}
    \caption{%
      Synthesis Algorithm for $\res$ semantics, with $S$ the set of
      states of $\ERG $, $\cc k_{\enrich{\beta}} \subseteq S$ the set of states
      with $\enrich \beta$-color $k$, and
      \textsc{Attractor}$(G, j, \cc k_{\enrich{\beta}})$ the attractor for
    \tpl{$j$} to~$\cc k_{\enrich{\beta}}$.\label{algo:zielonka}}
  \end{algorithm}
\end{minipage}
\hfill
\begin{minipage}{4.4cm}
  \begin{figure}[H]
    \centering
\begin{tikzpicture}[x=9mm,y=9mm]
  \draw[->] ( 0  ,-0.3) -- ( 3.5,-0.3) node[below] {$x$};
  \draw[->] (-0.3, 0  ) -- (-0.3, 3.5) node[left] {$y$};
  \foreach \pos in {0,1,2,3}
    {
      \draw[shift={(\pos,-0.3)}] (0pt,0pt) -- (0pt,-2pt) node[below] {$\pos$};
      \draw[shift={(-0.3,\pos)}] (0pt,0pt) -- (-2pt,0pt) node[left] {$\pos$};
    }

  \begin{scope}
    \clip (-0.1, -0.1) rectangle (3.45, 3.45);
    \foreach \x in {0,1,2,3}
    \foreach \y in {0,1,2,3}
      {
        \ifthenelse{\x=2 \AND \y=2}{\def\regcolor{TargetColor}}{
          \ifthenelse{\x=1 \AND \y=0}{\def\regcolor{BadColor}}{
            \ifthenelse{\x=\y \AND \x<2}{\def\regcolor{GoodColor}}{
              \def\regcolor{DisabledColor}
            }}}
        \FPeval\xx{round(\x+1:0)}
        \FPeval\yy{round(\y+1:0)}
        \def\regcolordown{\regcolor}
        \def\regcolorup{\regcolor}
        \ifthenelse{\x=\yy \AND \x=2}{\def\regcolorup{GoodColor}}{
          \ifthenelse{\x=\yy \AND \x<2}{\def\regcolorup{BadColor}}{
            \def\regcolorup{\regcolor}
          }}
        \ifthenelse{\y=\xx \AND \y<3}{\def\regcolordown{GoodColor}}{\def\regcolordown{\regcolor}}
        \ifthenelse{\x=0 \AND \y=0}{\def\regcolordown{DisabledColor}}{}
        \begin{scope}[shift={(\x, \y)},fill=\regcolor]
          \fill (-0.075, -0.075) rectangle (0.075, 0.075);
          \fill[fill=\regcolordown] (0.150, -0.075) -- (0.800, -0.075) -- (0.850, 0.075) -- (0.200, 0.075) -- cycle;
          \fill[fill=\regcolorup] (-0.075,  0.150) -- (-0.075, 0.800) -- (0.075, 0.850) -- (0.075, 0.200) -- cycle;
          \fill (0.150,  0.150) -- (0.150,  0.250) -- (0.750, 0.850) -- (0.850, 0.850) -- (0.850, 0.750) -- (0.250, 0.150) -- cycle;
          \fill[fill=\regcolordown] (0.360,  0.150) -- (0.850, 0.150) -- (0.850, 0.640) -- cycle;
          \fill[fill=\regcolorup] (0.150,  0.360) -- (0.150, 0.850) -- (0.640, 0.850) -- cycle;
        \end{scope}
      }
    \foreach \y in {-4,-3,...,4}
      {
        \draw[shift={(0, \y)},draw=white,line width=2.2pt] (-1, -0.845) -- (4, 4.155);
        \draw[shift={(0, \y)},draw=white,line width=2.2pt] (-0.845, -1) -- (4.155, 4);
      }
  \end{scope}
\end{tikzpicture}
    \caption{$\predt$ Valuations}
    \label{fig:predt}
  \end{figure}
  \begin{figure}[H]
\begin{tikzpicture}[auto, node distance=0.5cm and 1.8cm, align=center, initial text=,
    initial where=above,
    every label/.style={font=\tiny, inner sep=1pt, circle}]
  \node (qi) [state, label={below right:$1$}, initial] {$q_i$};
  \node (ql) [state, label={below right:$0$}, right = of qi] {$q_l$};
  \node (qh) [state, label={below right:$2$}, above = of ql] {$q_h$};
  \path [->, every node/.style={font=\footnotesize}]
  (qi) edge [thin      ] node [sloped, pos=0.3] {$h, x = 0$} (qh)
  (qi) edge [very thick] node [sloped, pos=0.65] {$l, x = 0$} (ql)
  (qh) edge [loop left, looseness=3.6] node {$h, x < 1$} (qh)
  ;
\end{tikzpicture}
    \caption{A \tga}
    \label{fig:subgame}
  \end{figure}
\end{minipage}

\paragraph*{Overview}

Algorithm~\ref{algo:zielonka} operates on the arena $\nameEtga$ and
relies on the coloring function $\enrich \beta$, as defined in
Section~\ref{sec:reduction-to-parity} (\wlogep, we assume that there
is at least one state of each color from $0$ to the maximum color
$d$). It manipulates infinite sets of states of $\ERG$ in the $S$,
$T_i$ and $W_i$ variables.  These set of states are symbolically
represented as finite sets of pairs
$(q, f) \in \enrich Q \times \mathbb F(X)$, i.e.\ $q$ is a location and
$f$ is a federation.

Like the Zielonka algorithm, \cref{algo:zielonka} solves a game by
recursively solving subgames.  These subgames are constructed so that
the winning strategy for one of the player in the subgame remains a
valid partial winning strategy in the current game.

The core components of Algorithm~\ref{algo:zielonka} are the
\textsc{Attractor} function (lines 4 and 12), and a difference
operator $\backslash_j$ (parameterized by a player number $j$) that
reduces the size of the arena before recursive calls (lines 8 and 13).
Intuitively, $\textsc{Attractor}(G, j, T)$ returns all the states from
which \tpl{$j$} can force game $G$ to reach the target set of states
$T$.

During each call, $j$ is the player with the highest color in the game
and $i$ is his adversary. Hence, initially $j=1$ iff $d$ is odd and
$i = 1-j$.  The attractor $T_j$ is computed for $j$ on the set of all
states of maximal color (line 4).  If all states of the arena are
winning (line 5), the algorithm terminates with $T_j = S = W_j$.
Otherwise, a subarena is computed by removing $T_j$ from the initial
arena $G$, and a recursive call is performed (line 8) to compute the
winning states \emph{for the other player}.  If no states are
discovered winning for him (line 9) the algorithm terminates with
$T_j = S = W_j$.  Otherwise, the attractor to these states $T_i$ is
removed and a recursive call is performed (line 13).  Since \tpl$i$'s
attractor $T_i$ was removed, we know that states winning for $j$ in
the subgame are exactly the states winning for $j$ in the current game
(line~14).
To initialize this process, the algorithm first computes $k$, the
largest $\enrich \beta$-color in $G$; $j$, the player for which $k$ is a good
color; and $i$, the other player. The target of the first attractor is
$\cc{k}_{\enrich \beta}$, defined as the set of all states of maximal
$\enrich \beta$ color.

In the end, given an initial location $q_\init$, the Controller has a
winning strategy $\res(G)$ from $q_\init$ whenever there is
$( (q_\init, \alpha(q_\init), \none), F ) \in W_0$ s.t.
$\mathbf 0_{X_z} \in \zsem{F}$.

\paragraph*{Attractor Computation}

\begin{definition}
  [Attractor] The \emph{attractor} for a \pp to a set of target states
  $T$ is the set of all states from which $p$ has a strategy that
  guarantees that the game evolves to a state of $T$ (including $T$).
  Formally, given a set of states $T$, a Controller's attractor to $T$
  is the set of all states $(q, v)$ s.t.\
  $\exists \sigma_0 : \forall \sigma_1 : \forall \rho \in \out{(q,
    v)}(\sigma_0, \sigma_1)$: $\rho$ visits $T$.

  An Environment's attractor to $T$ is defined accordingly. It is
  the set of all states $(q, v)$ s.t.\
  $\forall \sigma_0 : \exists \sigma_1 : \exists \rho \in \out{(q,
    v)}(\sigma_0, \sigma_1)$: $\rho$ visits $T$.
\end{definition}

Given an arena $G$, a \pp and a set of target states $T$, the call to
\textsc{Attractor}$(G, p, T)$ returns the attractor for $p$ to~$T$.
The efficient use of zones for the attractor computation relies on the
properties of semantics $\res$ emphasized
in~\ref{sec:algorithmic-properties}. In order to compute attractors,
we rely on the classical \emph{one-step controllable
  predecessors}. The \emph{one-step controllable predecessors} for a
\pp to a set of target states $T$ is the set of all states from which
$p$ will surely get into $T$ in one step, whatever the move of the
opponent, and is noted $\cpre_p(T)$. Assuming split strategies, the
computation of $\cpre_p(T)$ itself can be broken down into several steps that
we describe now.

\begin{description}
\item[$\predt$:] Given a game $\game$ with states $S$, the
  \emph{temporal predecessor} of a set of target states~$T$ avoiding a
  set of bad states $B$ is, for $p \in \{0, 1\}$:
  \( \predt_p(T, B) = (T \setminus B) \cup \{ s \in S \mid
  \exists \Delta \in \Rp \tst s \xrightharpoonup{(\Delta, \bot_p)} s',
  s' \in T \wedge (\nexists \Delta' \in \Rp \tst \Delta' \leq \Delta
  \wedge s \xrightharpoonup{(\Delta', \bot_{1-p})} s'' \wedge s'' \in
  B)
  \}
\).
For example, suppose we want to reach
$\{ (q, v) \mid v \models 2 \leq x < 3 \wedge 2 \leq y < 3 \} = T$ (in
green on \cref{fig:predt}), but $q$ has an Environment's outgoing edge
with guard $(1 \leq x < 2 \wedge y < 1)$ (in dark) going into a
location we want to avoid.  So, $B$ is
$\{(q, v') \mid v' \models 1 \leq x < 2 \wedge y < 1 \}$.  Then
$\predt_0(T, B)$ is $\{ (s, v'') \mid v'' \in f \}$ with $f$ the set
of valuations in yellow and green on the figure.

$\predt_p(T, B)$ can be computed by manipulating federations of zones.
Given a target set~$T$, let
$\tbefore T=\{ (q, \tbefore f) \mid (q, f) \in T \}$.  As shown
in~\cite{FCassezDavidFleuryLarsenLime05EfficientFlyAlgorithms.a},
$\predt_p(T, B) = (T \setminus B) \cup (\tbefore T \setminus \tbefore
B)$ where
$T \setminus B = \{ (q, f) \in T \mid \nexists (q, f') \in B \} \cup
\{ (q, f \setminus f') \in T\in T \mid \exists (q, f') \in B \}$.

\item[$\dcpre_0$:] The Controller's discrete controllable predecessors
  of a set $T$ is the set of states $s$ s.t.:
\(
  \dcpre_0(T) = \{s \in S \mid 
  (\exists a_0 \in A_0 : s \xrightharpoonup{0, a_0} s', s' \in T
  \wedge \nexists a_1 \in A_1 : s \xrightharpoonup{0, a_1} s'', s'' \not\in T) 
  \vee
  (\neg P_0(s)
  \wedge \exists a_1 \in A_1 : s \xrightharpoonup{0, a_1} s'', s'' \in T
  \wedge \nexists a_1' \in A_1 : s \xrightharpoonup{0, a_1'} s'', s'' \not\in T) \}
\).
The second part of the disjunction captures the case where the Controller does not play in the $\res$ semantics.

\item[$\dcpre_1$:] The Environment's discrete controllable predecessor of a set of target states $T$ is the set of states $s$ s.t.:
\(
  \dcpre_1(T) = \{ s \in S \mid 
    \exists a_1 \in A_1 : s \xrightharpoonup{0, a_1} s', s' \in T
  \vee
\forall a_0 \in A_0 : s \xrightharpoonup{0, a_0} s'', s'' \in T \}
\).

\item[$\cpre_p$:] Finally, \pp's controllable predecessor of a set of
  target states $T$ can be computed as:
  \( \cpre_p(T) = \predt_p(\dcpre_p(T), \dcpre_{1-p}(S \setminus T))
  \).
\end{description}
Now that we have the computation of $\cpre_p(T)$, we can use a
standard fixed point algorithm to obtain the attractor:
\begin{lemma}
  The attractor for a \pp to a set of states~$T$ is the closure $\left(\cpre_p\right)^*(T)$.
\end{lemma}

\paragraph*{Subgame Creation}
Let us finally explain how we compute subgames with the $\setminus_i$
operator. Let us first explain why this requires careful consideration
through the example in \cref{fig:subgame}. Here, the Controller wins
by reaching either $q_h$ or $q_l$.  The Controller wins this game by
playing $(0, h)$ in $q_i$: either this move gets played and $q_h$ is
reached, or the Environment plays $(0, l)$ and reaches $q_l$.

The algorithm starts by computing the attractor of the locations with
maximal colors, i.e.\ $\textsc{Attractor}\big(G,0,(q_h, x=1)\big)$.
This attractor is $(q_h, 0\leq x\leq 1)$ and does not contain
$(q_i, x=0)$ since the Environment can force the transition from $q_i$
to $q_l$ in this state and thus avoid $q_h$.  Next, the algorithm
needs to solve a subgame from which we have removed at least the
attractor $(q_h, 0\leq x\leq 1)$. However, this example shows we need
to remove more than simply location $q_h$ and the transition from
$q_i$ to $q_h$.  Indeed, in the same game without $q_h$, the
Environment wins by waiting a non-null delay in $q_i$, which prevents
reaching $q_l$ (although this strategy does not make him win in the
full game).  This shows the difficulty: we want to remove transitions
and states from the transition system while the algorithm works on the
arena.  Our difference operator is defined to take this into
account.

Concretely, the $\setminus_i$ operator does not only remove locations
and transitions, but also modifies the no-wait functions $\urg p$ to
disable continuous transitions.  Thus, let us first refine the
$\predt$ operator to account for the $U$ functions: Every time we
write $\predt_p(T, B)$, we implicitly mean
$\predt_p(T, B \cup \overline U)$ with
$\overline U = \{(q, \urg p(q)) \mid q \in Q)\} \setminus T$.

Given an attractor $T_0$ of \tpl0, let $T'$ denote the set of states
$s$ not in $T_0$ where there exists an enabled discrete transition of
the Controller from $s$ to some state in $T_0$.  (So we know that
there is an enabled Environment's discrete transition from $s$ to a
state out of $T_0$.)  Then we define:
$G \setminus_0 T_0 := (Q, E, C', \urg0, \urg1, \alpha)$ with
$C'(q) = C(q) \setminus f$ if $\exists (q, f) \in T_0$ else $C(q)$,
$\urg1'(q) = \urg1(q) \setminus f$ if $\exists (q, f) \in T'$ else
$\urg1(q)$.  Removing a \tpl1's attractor is straightforward.  Given
an arena $\nametga$ and a set of states $T_1$ that is an attractor for
\tpl1, we define:
$G \setminus_1 T_1 := (Q, E, C', \urg0, \urg1, \alpha)$ with
$C'(q) = C(q) \setminus f$ if $\exists (q, f) \in T_1$ else $C(q)$.

We close this section by sketching the main ideas behind the proof of
correctness of the algorithm. 

\begin{theorem}\label{theo:algo}
 \Cref{algo:zielonka} terminates and returns the winning sets of both players.
\end{theorem}

Termination is established by showing that the depth of recursive
calls is bounded, using the classical region construction. The set of
regions $Q \times \cregs(G)$ of a game $\res(G)$ is finite.  By
definition, the attractor $\textsc{Attractor}(G,j,T)$ never returns an
empty set, since it contains $T$.  Thus, each recursive call is
performed on a game with at least one region less than the game in the
calling procedure, which establishes termination.

The correctness is established inductively. The \textbf{induction hypothesis}
is that \textsc{Solve$(G)$} returns a partition $(W_0, W_1)$ of the
states of $\ERG$ s.t.\ $p$ has a winning strategy from all states
of $W_p$.

To prove the inductive step, we consider all the return
statements. The partition returned on line 7 satisfies the induction
hypothesis by properties of the attractor.  On line 11, we have a
\tpl$j$ attractor $T_j$ and a $j$-winning strategy from every state in
the subgame $G \setminus_j T_j$.  For the final return statement, note
that $T_i$ is an $i$-attractor to a set $W'_i$ constructed s.t.\ $i$
has a winning strategy from $W'_i$ in $G$.  Thus, $T_i$ is winning for
$i$.  Finally, from the states of $G \setminus_i T_i$, if a player has
a winning strategy in this subgame, then they also have a winning
strategy in $G$. To prove that the return on line 15 is correct, we
rely on the properties of our dedicated difference operator
$\setminus_i$. 

\subsection{Efficient zone-based computation thanks to the \texorpdfstring{$\res$}{ℛ} semantics}\label{sec:algorithmic-properties}

In this section, we explain why the definition of the $\res$ semantics
is essential to obtain efficient zone-based algorithms. As mentioned
before, the main takeaway message is that the $\res$ semantics (with
the lazy blame policy) allows avoiding splitting zones into too many
regions when computing the predecessor operator.

This problem is illustrated in \cref{fig:cpre}. First, remember that,
in $\enrich G$, the states contain an extra bit indicating which
player is to blame. This \emph{discrete} information present in the states
will be the source of the difficulties we are about to discuss. Assume
we have game with a single location $q$, and assume we have obtained
the symbolic state $(q,1,T)$, where $1$ indicates that the environment
has the blame and $T$ is the zone depicted in
\cref{fig:cpre-t}. First, observe that
$\cpre_0(\{(q,1,T)\})=\{(q,1,T)\}$. Indeed, the Controller cannot
force, in one step, to reach $(q,1,T)$ from any other state, because
the Controller has never the guarantee to play: either the Environment
plays faster, or the non-determinism is resolved in favour of the
Environment, who takes the blame (while letting the Controller play
would have changed the blame bit to $0$). Next, let us consider the
computation of the Environment's Controllable Predecessors
$\cpre_1(\{(q,1,T)\})$. In \cref{fig:cpre-bad}, we can observe that,
if we had applied the ``classical'' blame policy to the $\res$
semantics, $\cpre_1(\{(q,1,T)\})$ would have a non-convex form, which
means that it needs to be represented by a federation containing at least
three zones. For example, region $x=2\wedge 1<y<2$
is not contained in $\cpre_1(\{(q,1,T)\})$ because the Controller
could play a shorter delay than the one proposed by the Environment
and take the blame, flipping the blame bit from $1$ to $0$.
Using our lazy blame rule \emph{that does not blame long waits}
however preserves the convexity of $\cpre_1(\{(q,1,T)\})$, as seen in
\cref{fig:cpre-good}. The lazy blame avoid flipping the blame bit in
the states and allows one to compute larger convex sets at once with
the $\cpre$ operator.
\setlength{\tmplen}{0.33\textwidth-1.2ex}
\begin{figure}[ht]
  \begin{subfigure}[t]{\tmplen}
    \centering
\begin{tikzpicture}[x=7mm,y=7mm]
  \draw[->] ( 0  ,-0.3) -- ( 3.5,-0.3) node[below] {$x$};
  \draw[->, overlay] (-0.3, 0  ) -- (-0.3, 3.5) node[left] {$y$};
  \foreach \pos in {0,1,2,3}
  {
    \draw[shift={(\pos,-0.3)}] (0pt,0pt) -- (0pt,-2pt) node[below] {$\pos$};
    \draw[shift={(-0.3,\pos)}, overlay] (0pt,0pt) -- (-2pt,0pt) node[left] {$\pos$};
  }

  \begin{scope}
  \clip (-0.1, -0.1) rectangle (3.45, 3.45);
  \foreach \x in {0,1,2,3}
    \foreach \y in {0,1,2,3}
    {
      \ifthenelse{\x=2 \AND \y=2}{\def\regcolor{TargetColor}}{\def\regcolor{DisabledColor}}
      \def\regcolorup{\regcolor}
      \def\regcolordown{\regcolor}
      \begin{scope}[shift={(\x, \y)},fill=\regcolor]
        \fill (-0.075, -0.075) rectangle (0.075, 0.075);
        \fill[fill=\regcolordown] ( 0.150, -0.075) -- (0.800, -0.075) -- (0.850, 0.075) -- (0.200, 0.075) -- cycle;
        \fill[fill=\regcolorup] (-0.075,  0.150) -- (-0.075, 0.800) -- (0.075, 0.850) -- (0.075, 0.200) -- cycle;
        \fill ( 0.150,  0.150) -- ( 0.150,  0.250) -- (0.750, 0.850) -- (0.850, 0.850) -- (0.850, 0.750) -- (0.250, 0.150) -- cycle;
        \fill[fill=\regcolordown] ( 0.360,  0.150) -- (0.850, 0.150) -- (0.850, 0.640) -- cycle;
        \fill[fill=\regcolorup] ( 0.150,  0.360) -- (0.150, 0.850) -- (0.640, 0.850) -- cycle;
      \end{scope}
    }
  \foreach \y in {-4,-3,...,4}
  {
    \draw[shift={(0, \y)},draw=white,line width=2.2pt] (-1, -0.845) -- (4, 4.155);
    \draw[shift={(0, \y)},draw=white,line width=2.2pt] (-0.845, -1) -- (4.155, 4);
  }
  \end{scope}
\end{tikzpicture}
    \caption{A zone $T$.
    With the classic blame, we have $\{(q,1,T)\} = \cpre_0(\{(q,1,T)\})$.}\label{fig:cpre-t}
  \end{subfigure}
  \hspace*{0.6ex}
  \begin{subfigure}[t]{\tmplen}
    \centering
\begin{tikzpicture}[x=7mm,y=7mm]
  \draw[->] ( 0  ,-0.3) -- ( 3.5,-0.3) node[below] {$x$};
  \foreach \pos in {0,1,2,3}
  {
    \draw[shift={(\pos,-0.3)}] (0pt,0pt) -- (0pt,-2pt) node[below] {$\pos$};
  }

  \begin{scope}
  \clip (-0.1, -0.1) rectangle (3.45, 3.45);
  \foreach \x in {0,1,2,3}
    \foreach \y in {0,1,2,3}
    {
      \ifthenelse{\x=2 \AND \y=2}{\def\regcolor{TargetColor}}{\def\regcolor{DisabledColor}}
      \ifthenelse{\x=1 \AND \y=1}{\def\regcolormid{GoodColor}}{\def\regcolormid{\regcolor}}
      \ifthenelse{\x=2 \AND \y=1}{\def\regcolorup{GoodColor}}{\def\regcolorup{\regcolor}}
      \ifthenelse{\x=1 \AND \y=2}{\def\regcolordown{GoodColor}}{\def\regcolordown{\regcolor}}
      \begin{scope}[shift={(\x, \y)},fill=\regcolor]
        \fill (-0.075, -0.075) rectangle (0.075, 0.075);
        \fill ( 0.150, -0.075) -- (0.800, -0.075) -- (0.850, 0.075) -- (0.200, 0.075) -- cycle;
        \fill (-0.075,  0.150) -- (-0.075, 0.800) -- (0.075, 0.850) -- (0.075, 0.200) -- cycle;
        \fill[fill=\regcolormid] ( 0.150,  0.150) -- ( 0.150,  0.250) -- (0.750, 0.850) -- (0.850, 0.850) -- (0.850, 0.750) -- (0.250, 0.150) -- cycle;
        \fill[fill=\regcolordown] ( 0.360,  0.150) -- (0.850, 0.150) -- (0.850, 0.640) -- cycle;
        \fill[fill=\regcolorup] ( 0.150,  0.360) -- (0.150, 0.850) -- (0.640, 0.850) -- cycle;
      \end{scope}
    }
  \foreach \y in {-4,-3,...,4}
  {
    \draw[shift={(0, \y)},draw=white,line width=2.2pt] (-1, -0.845) -- (4, 4.155);
    \draw[shift={(0, \y)},draw=white,line width=2.2pt] (-0.845, -1) -- (4.155, 4);
  }
  \end{scope}
\end{tikzpicture}
    \caption{A federation $T'' \supset T$.
      With the classic blame, $\cpre_1(\{(q,1,T)\}) = \{(q, 1, T'')\}$.}\label{fig:cpre-bad}
  \end{subfigure}
  \hspace*{0.6ex}
  \begin{subfigure}[t]{\tmplen}
    \centering
\begin{tikzpicture}[x=7mm,y=7mm]
  \draw[->] ( 0  ,-0.3) -- ( 3.5,-0.3) node[below] {$x$};
  \foreach \pos in {0,1,2,3}
  {
    \draw[shift={(\pos,-0.3)}] (0pt,0pt) -- (0pt,-2pt) node[below] {$\pos$};
  }

  \begin{scope}
  \clip (-0.1, -0.1) rectangle (3.45, 3.45);
  \foreach \x in {0,1,2,3}
    \foreach \y in {0,1,2,3}
    {
      \ifthenelse{\x=2 \AND \y=2}{\def\regcolor{TargetColor}}{
        \ifthenelse{\x=\y \AND \x<2}{\def\regcolor{GoodColor}}{\def\regcolor{DisabledColor}}
      }
      \FPeval\xx{round(\x+1:0)}
      \FPeval\yy{round(\y+1:0)}
      \ifthenelse{\x=\yy \AND \x<3}{\def\regcolorup{GoodColor}}{\def\regcolorup{\regcolor}}
      \ifthenelse{\y=\xx \AND \y<3}{\def\regcolordown{GoodColor}}{\def\regcolordown{\regcolor}}
      \begin{scope}[shift={(\x, \y)},fill=\regcolor]
        \fill (-0.075, -0.075) rectangle (0.075, 0.075);
        \fill[fill=\regcolordown] (0.150, -0.075) -- (0.800, -0.075) -- (0.850, 0.075) -- (0.200, 0.075) -- cycle;
        \fill[fill=\regcolorup] (-0.075,  0.150) -- (-0.075, 0.800) -- (0.075, 0.850) -- (0.075, 0.200) -- cycle;
        \fill (0.150,  0.150) -- (0.150,  0.250) -- (0.750, 0.850) -- (0.850, 0.850) -- (0.850, 0.750) -- (0.250, 0.150) -- cycle;
        \fill[fill=\regcolordown] (0.360,  0.150) -- (0.850, 0.150) -- (0.850, 0.640) -- cycle;
        \fill[fill=\regcolorup] (0.150,  0.360) -- (0.150, 0.850) -- (0.640, 0.850) -- cycle;
      \end{scope}
    }
  \foreach \y in {-4,-3,...,4}
  {
    \draw[shift={(0, \y)},draw=white,line width=2.2pt] (-1, -0.845) -- (4, 4.155);
    \draw[shift={(0, \y)},draw=white,line width=2.2pt] (-0.845, -1) -- (4.155, 4);
  }
  \end{scope}
\end{tikzpicture}
    \caption{A zone $T' \supset T$.
      With the lazy blame, $\cpre_0(\{(q,1,T)\}) = \cpre_1(\{(q,1,T)\}) = \{(q, 1, T')\}$}\label{fig:cpre-good}
  \end{subfigure}
  \caption{The lazy blame simplifies the controllable predecessor
    computation.}\label{fig:cpre}
\end{figure}
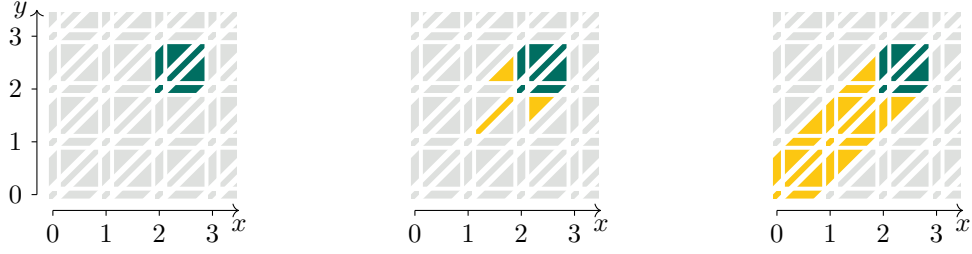

\section{Empirical Results}

We have a proof of concept implementation which uses \uppaal's federation library.
We have tested it with artificial examples of varying sizes.
We focus on time, as memory usage has remained relatively low in our experiments. 
These results aim to provide a general idea of the algorithm's behavior rather than measure its effectiveness precisely.

Here are the main takeaways.
As expected, it exhibits exponential time growth with respect to the number of clocks, polynomial time growth with respect to the number of distinct guards.
We can observe that the algorithm effectively exploits the zone data-structure to capture the structure of the game.
The problematic behavior explained in \cref{fig:cpre-bad} has not been observed.
We now focus on an example showing this last point.

The family of arenas informally depicted on \cref{fig:test-1} shows that the algorithm effectively harness the zone data-structures to capture the structure of the game.
The arena is parametrized by an integer $n\geq1$.
It has $3n + 1$ locations and $3n(n+1)$ edges, one clock $x$ with $c_x = n$ and the largest color is 3.
The locations are $t$ and $\big\{a_i, b_i, s_i\big\}_{i\in\{1, \dots, n\}}$.
The Environment edges are
$\{(a_i, e_b, \top, \emptyset, b_i),\ \allowbreak
  (b_i, e_b, \top, \emptyset, a_i),\ \allowbreak
  (a_i, e_{a_j}, \top, \{x\}, a_j),\ \allowbreak
  (a_i, e_{s_i}, \top, \{x\}, s_i),\ \allowbreak
  (a_i, e_{s_j}, \top, \{x\}, s_j),\ \allowbreak
  (s_i, e_{s_i}, \top, \{x\}, s_i),\ \allowbreak
  (s_i, e_{s_j}, \top, \{x\}, s_j)
\mid \forall i,j \in \{1, …, n\}, j \neq i \}$.
Controller edges are
$\{(s_i, c_t, 0<x<1, \emptyset, t),\ \allowbreak
  (t, c_{a_i}, \top, \{x\}, a_i)
\mid \forall i \in \{1, …, n\} \}$.
In this game, the number of locations is growing linearly with $n$ and the number of edges quadratically.

\begin{figure}
  \centering
\begin{tikzpicture}[auto, node distance=0.9cm and 1.8cm, align=center, initial text=,
    every label/.style={font=\tiny, inner sep=1pt, circle}]
  \node (ai) [state, label={below right:$0$}] {$a_i$} ;
  \node (bi) [state, label={below right:$1$}, left  = of ai] {$b_i$} ;
  \node (si) [state, label={below right:$2$}, above = of ai] {$s_i$} ;
  \node (t)  [state, label={below right:$3$}, right = of si] {$t$} ;
  \node (sj) [left = of si] {$s_{j\neq i}$} ;
  \node (aj) [right = of ai] {$a_{j\neq i}$} ;
  \node[below left=0.3cm and 0.3cm of ai] (init) {};
  \draw[->] (init) -- (ai);
  \node [style={font=\footnotesize}, above] at (t.north) {$x<n$};
  \node [style={font=\footnotesize}, below] at (ai.south) {$x<1$};
  \path [->, every node/.style={font=\footnotesize}, very thick]
  (ai) edge [bend left=20] (bi)
  (bi) edge [bend left=20] (ai)
  (ai) edge node [swap] {$x:=0$} (si)
  (ai) edge node [sloped] {$x:=0$} (sj)
  (ai) edge (aj)
  (si) edge [bend right] node [swap] {$x:=0$} (sj)
  (si) edge [loop left, looseness=3.6] node {$x:=0$} (si)
  ;
  \path [->, every node/.style={font=\footnotesize}, thin]
  (si) edge node {$0<x<1$} (t)
  (t)  edge node [sloped, swap] {$x:=0$} (ai)
  (t)  edge node {$x:=0$} (aj)
  ;
\end{tikzpicture}
  \caption{First Test}
  \label{fig:test-1}
\end{figure}

%
The calculation time is linear in the number of edges, showing that the zone data-structure effectively capture the structure of the game.
A size limitation on the zones, as in \cref{fig:cpre-bad}, would have resulted in an exponential blow-up.
We tested two variations: one with both invariants as $x < 1$ and another with both as $x < n$.
The $x < n$ case yields faster computation than the depicted case.
The $x < 1$ case is slower.
Both maintain linearity, supporting the intuition.
For these arenas, the number of recursive calls does not depend on the size.
The depicted game involves 37 calls.
The $x < 1$ case requires 28 calls, while $x < n$ demands 45.

\clearpage

\bibliography{references}

\begin{thebibliography}{10}

\bibitem{RAlurDill94TheoryTimedAutomata.a}
Rajeev Alur and David~L. Dill.
\newblock A theory of timed automata.
\newblock {\em Theor. Comput. Sci.}, 126(2):183–235, 1994.
\newblock \href {https://doi.org/10.1016/0304-3975(94)90010-8}
  {\path{doi:10.1016/0304-3975(94)90010-8}}.

\bibitem{DBLP:conf/hybrid/AsarinMP94}
Eugene Asarin, Oded Maler, and Amir Pnueli.
\newblock Symbolic controller synthesis for discrete and timed systems.
\newblock In Panos~J. Antsaklis, Wolf Kohn, Anil Nerode, and Shankar Sastry,
  editors, {\em Hybrid Systems II, Proceedings of the Third International
  Workshop on Hybrid Systems, Ithaca, NY, USA, October 1994}, volume 999 of
  {\em Lecture Notes in Computer Science}, page 1–20. Springer, 1994.
\newblock \href {https://doi.org/10.1007/3-540-60472-3\_1}
  {\path{doi:10.1007/3-540-60472-3\_1}}.

\bibitem{DBLP:conf/concur/BouyerJM15}
Patricia Bouyer, Samy Jaziri, and Nicolas Markey.
\newblock On the value problem in weighted timed games.
\newblock In Luca Aceto and David de~Frutos{-}Escrig, editors, {\em 26th
  International Conference on Concurrency Theory, {CONCUR} 2015, Madrid, Spain,
  September 1.4, 2015}, volume~42 of {\em LIPIcs}, page 311–324. Schloss
  Dagstuhl - Leibniz-Zentrum für Informatik, 2015.
\newblock URL: \url{https://doi.org/10.4230/LIPIcs.CONCUR.2015.311}, \href
  {https://doi.org/10.4230/LIPICS.CONCUR.2015.311}
  {\path{doi:10.4230/LIPICS.CONCUR.2015.311}}.

\bibitem{DBLP:conf/formats/BrihayeBR05}
Thomas Brihaye, Véronique Bruyère, and Jean{-}François Raskin.
\newblock On optimal timed strategies.
\newblock In Paul Pettersson and Wang Yi, editors, {\em Formal Modeling and
  Analysis of Timed Systems, Third International Conference, {FORMATS} 2005,
  Uppsala, Sweden, September 26-28, 2005, Proceedings}, volume 3829 of {\em
  Lecture Notes in Computer Science}, page 49–64. Springer, 2005.
\newblock \href {https://doi.org/10.1007/11603009\_5}
  {\path{doi:10.1007/11603009\_5}}.

\bibitem{DBLP:conf/icalp/BrihayeHPR07}
Thomas Brihaye, Thomas~A. Henzinger, Vinayak~S. Prabhu, and Jean{-}François
  Raskin.
\newblock Minimum-time reachability in timed games.
\newblock In Lars Arge, Christian Cachin, Tomasz Jurdzinski, and Andrzej
  Tarlecki, editors, {\em Automata, Languages and Programming, 34th
  International Colloquium, {ICALP} 2007, Wroclaw, Poland, July 9-13, 2007,
  Proceedings}, volume 4596 of {\em Lecture Notes in Computer Science}, page
  825–837. Springer, 2007.
\newblock \href {https://doi.org/10.1007/978-3-540-73420-8\_71}
  {\path{doi:10.1007/978-3-540-73420-8\_71}}.

\bibitem{FCassezDavidFleuryLarsenLime05EfficientFlyAlgorithms.a}
Franck Cassez, Alexandre David, Emmanuel Fleury, Kim~Guldstrand Larsen, and
  Didier Lime.
\newblock Efficient on-the-fly algorithms for the analysis of timed games.
\newblock In Martín Abadi and Luca de~Alfaro, editors, {\em {CONCUR} 2005 -
  Concurrency Theory, 16th International Conference, {CONCUR} 2005, San
  Francisco, CA, USA, August 23-26, 2005, Proceedings}, volume 3653 of {\em
  Lecture Notes in Computer Science}, page 66–80. Springer, 2005.
\newblock \href {https://doi.org/10.1007/11539452\_9}
  {\path{doi:10.1007/11539452\_9}}.

\bibitem{DBLP:conf/hybrid/CassezHR02}
Franck Cassez, Thomas~A. Henzinger, and Jean{-}François Raskin.
\newblock A comparison of control problems for timed and hybrid systems.
\newblock In Claire~J. Tomlin and Mark~R. Greenstreet, editors, {\em Hybrid
  Systems: Computation and Control, 5th International Workshop, {HSCC} 2002,
  Stanford, CA, USA, March 25-27, 2002, Proceedings}, volume 2289 of {\em
  Lecture Notes in Computer Science}, page 134–148. Springer, 2002.
\newblock \href {https://doi.org/10.1007/3-540-45873-5\_13}
  {\path{doi:10.1007/3-540-45873-5\_13}}.

\bibitem{DBLP:conf/hybrid/CassezJLRR09}
Franck Cassez, Jan~Jakob Jessen, Kim~Guldstrand Larsen, Jean{-}François
  Raskin, and Pierre{-}Alain Reynier.
\newblock Automatic synthesis of robust and optimal controllers - an industrial
  case study.
\newblock In Rupak Majumdar and Paulo Tabuada, editors, {\em Hybrid Systems:
  Computation and Control, 12th International Conference, {HSCC} 2009, San
  Francisco, CA, USA, April 13-15, 2009. Proceedings}, volume 5469 of {\em
  Lecture Notes in Computer Science}, page 90–104. Springer, 2009.
\newblock \href {https://doi.org/10.1007/978-3-642-00602-9\_7}
  {\path{doi:10.1007/978-3-642-00602-9\_7}}.

\bibitem{LAlfaroFaellaHenzingerMajumdarStoelinga03ElementSurpriseTimed.a}
Luca de~Alfaro, Marco Faella, Thomas~A. Henzinger, Rupak Majumdar, and
  Mariëlle Stoelinga.
\newblock The element of surprise in timed games.
\newblock In Roberto~M. Amadio and Denis Lugiez, editors, {\em {CONCUR} 2003 -
  Concurrency Theory, 14\textsuperscript{th} International Conference,
  Marseille, France, September 3-5, 2003, Proceedings}, volume 2761 of {\em
  Lecture Notes in Computer Science}, page 142–156. Springer, 2003.
\newblock \href {https://doi.org/10.1007/978-3-540-45187-7\_9}
  {\path{doi:10.1007/978-3-540-45187-7\_9}}.

\bibitem{DBLP:conf/formats/GiampaoloGRS10}
Barbara {Di Giampaolo}, Gilles Geeraerts, Jean{-}François Raskin, and Nathalie
  Sznajder.
\newblock Safraless procedures for timed specifications.
\newblock In Krishnendu Chatterjee and Thomas~A. Henzinger, editors, {\em
  Formal Modeling and Analysis of Timed Systems - 8th International Conference,
  {FORMATS} 2010, Klosterneuburg, Austria, September 8-10, 2010. Proceedings},
  volume 6246 of {\em Lecture Notes in Computer Science}, page 2–22.
  Springer, 2010.
\newblock \href {https://doi.org/10.1007/978-3-642-15297-9\_2}
  {\path{doi:10.1007/978-3-642-15297-9\_2}}.

\bibitem{DBLP:conf/formats/DoyenGRR09}
Laurent Doyen, Gilles Geeraerts, Jean{-}François Raskin, and Julien Reichert.
\newblock Realizability of real-time logics.
\newblock In Joël Ouaknine and Frits~W. Vaandrager, editors, {\em Formal
  Modeling and Analysis of Timed Systems, 7th International Conference,
  {FORMATS} 2009, Budapest, Hungary, September 14-16, 2009. Proceedings},
  volume 5813 of {\em Lecture Notes in Computer Science}, page 133–148.
  Springer, 2009.
\newblock \href {https://doi.org/10.1007/978-3-642-04368-0\_12}
  {\path{doi:10.1007/978-3-642-04368-0\_12}}.

\bibitem{DBLP:conf/rtss/LarsenLPY97}
Kim~Guldstrand Larsen, Fredrik Larsson, Paul Pettersson, and Wang Yi.
\newblock Efficient verification of real-time systems: compact data structure
  and state-space reduction.
\newblock In {\em Proceedings of the 18th {IEEE} Real-Time Systems Symposium
  {(RTSS} '97), December 3-5, 1997, San Francisco, CA, {USA}}, page 14–24.
  {IEEE} Computer Society, 1997.
\newblock \href {https://doi.org/10.1109/REAL.1997.641265}
  {\path{doi:10.1109/REAL.1997.641265}}.

\bibitem{DBLP:journals/fac/WulfDR05}
Martin~De Wulf, Laurent Doyen, and Jean{-}François Raskin.
\newblock Almost {ASAP} semantics: from timed models to timed implementations.
\newblock {\em Formal Aspects Comput.}, 17(3):319–341, 2005.
\newblock URL: \url{https://doi.org/10.1007/s00165-005-0067-8}, \href
  {https://doi.org/10.1007/S00165-005-0067-8}
  {\path{doi:10.1007/S00165-005-0067-8}}.

\bibitem{WZielonka98InfiniteGamesFinitely.a}
Wieslaw Zielonka.
\newblock Infinite games on finitely coloured graphs with applications to
  automata on infinite trees.
\newblock {\em Theoretical Computer Science}, 200(1):135–183, 1998.
\newblock URL:
  \url{https://www.sciencedirect.com/science/article/pii/S0304397598000097},
  \href {https://doi.org/10.1016/S0304-3975(98)00009-7}
  {\path{doi:10.1016/S0304-3975(98)00009-7}}.

\end{thebibliography}

\end{document}